\documentclass[onecolumn]{aa}
\usepackage{graphicx}
\usepackage{multirow}
\usepackage{txfonts}

\def\lsim{\lower.5ex\hbox{$\; \buildrel < \over \sim \;$}}
\def\gsim{\lower.5ex\hbox{$\; \buildrel > \over \sim \;$}}

\begin{document}
\title{Accretion flow dynamics during the evolution of timing and spectral properties of GX 339-4 during its 2010-11 outburst}
\author{Anuj Nandi\inst{1,2}, Dipak Debnath\inst{2}, Samir Mandal\inst{3} and Sandip K. Chakrabarti\inst{4,2}} 
\offprints{Prof. Sandip K. Chakrabarti}
\institute{Space Astronomy Group, ISRO Satellite Centre, HAL Airport Road, Bangalore 560017, India 
\and
Indian Centre for Space Physics, Chalantika 43, Garia Station Rd., Kolkata, 700084, India 
\and
Indian Institute of Space Science and Technology, Thiruvananthapuram, 695547, India 
\and
S. N. Bose National Centre for Basic Sciences, Salt Lake, Kolkata 700098, India \\
   \email{anuj@isac.gov.in; dipak@csp.res.in; samir@iist.ac.in; chakraba@bose.res.in}}
   \date{Received 7 August 2011; accepted 28 March 2012}

\abstract
%% \\\\\context\\\\\
{The Galactic transient black-hole candidate GX 339-4 exhibited several outbursts at regular 
intervals of about two to three years in the Rossi X-ray Timing Explorer (RXTE) era. After 
remaining in an almost quiescent state for three long years, it again became X-ray active in 
January, 2010, continuing to be so over the next $\sim 14$ months.}
%% \\\\\aims\\\\\
{We study the timing and spectral properties of the black hole candidate (BHC) during its
recent outburst and understand the behavioral change in the accretion flow dynamics associated 
with the evolution of the various X-ray features.}
%% heading (mandatory) \\\\\methods\\\\\\\
{The detailed analysis of the temporal and spectral properties of the source during this outburst 
are carried out using archival data of the RXTE PCA instrument.We analyze a total of 
$236$ observational intervals consisting of $419$ days of data observed by RXTE,
from 2010 January 12 to 2011 March 6.}
%% heading (mandatory) \\\\\\results\\\\\\
{Our study provides a comprehensive understanding of the mass accretion processes and properties of 
the accretion disk of the black hole candidate. The PCA spectra of $2.5-25$ keV are 
mainly fitted with a combination of two components, namely, a disk black body and a power-law. 
The entire outburst as observed by RXTE, is divided into four spectral states, namely, 
hard, hard-intermediate, soft-intermediate, and soft. 
Quasi-periodic oscillations (QPOs) were found in three out of the four states, 
namely hard, hard-intermediate, and soft-intermediate states. The QPO frequencies increase 
monotonically from $0.102$ Hz to $5.692$ Hz in the rising phase
of the outburst, while during the declining phase QPO frequencies decrease
monotonically from $6.420$ to $1.149$ Hz. 
The evolution pattern, i.e. the hardness-intensity diagram, of the present outburst can be 
reproduced by two different components of the flow of accreting material.}
%%\\\\\\Conclusion\\\\\\\
{The recent outburst of GX 339-4 gives us an opportunity to understand the evolution of the 
two-component accretion rates starting from the onset to the end of the outburst phase. 
We found that the QPO frequency variation could be explained by the propagating oscillatory shock 
model (POS) and the hardness versus intensity variation can be reproduced if we assume that higher 
viscosity causes the conversion of a low angular momentum disk component into a Keplerian component 
during the outburst phase. The decline phase starts because of the reduction in the viscosity.}

\titlerunning{Spectral and Timing evolutions of GX 339-4 during its 2010-11 outburst}
\authorrunning{Nandi, Debnath, Mandal, and Chakrabarti}
\maketitle

\section{Introduction}

The Galactic transient black hole (BH) candidates are the most interesting X-ray objects to study, 
as these sources undergo peculiar temporal and spectral changes during their outburst phases.
Several spectral states are identified in BH outburst sources based on the evolution of their 
spectral and timing properties (McClintock \& Remillard 2003; Belloni et al. 2005; 
Remillard \& McClintock 2006). The spectral evolution of states in BH transients are also found to be 
associated with different branches of a q-like pattern (hysteresis loop), which is an X-ray 
hardness-intensity diagram (HID) (Maccarone \& Coppi 2003; Homan \& Belloni 2005) in the log-log plot. 
The existence of the similar loop feature is also reported for the neutron star binaries 
(Maccarone \& Coppi 2003) as well as for the CVs (Kording et al. 2008). In general, during an 
outburst, the spectral states are identified as four basic states -
{\it hard, hard-intermediate, soft-intermediate}, and {\it soft} states. One can find 
extensive discussions in the literature of the basic properties of all these four states 
(Homan \& Belloni 2005; Debnath et al. 2008; Motta et al. 2009; Dunn et al. 2010). 
In addition, a different state, termed the {\it very high} state, is also observed 
corresponding to the highest luminosity of some of the outburst sources 
(see Miyamoto et al. 1991; Remillard et al. 1999).
 
The temporal analysis of the X-ray variabilities observed in the outburst sources also reveals a
unique type of evolution in the power density spectrum (PDS). The PDSs are composed of different 
type of components, such as a flat-top (white noise) component, a break frequency (zero-centered 
Lorentzian function), and a power-law like distribution (red noise) etc. with a low-to-intermediate 
quasi-periodic oscillations (QPOs) ($<$ 20 Hz) (Belloni et al. 2002; Titarchuk et al. 2007). 
The observed QPOs are classified into three different classes `A', `B', and `C' types (see for details, 
Wijnands et al. 1999; Homan et al. 2001; Remillard et al. 2002), mostly based on the coherence 
factor Q (= $\nu/\Delta \nu$, where $\nu$ is the centroid frequency of QPO and $\Delta \nu$ is the FWHM) 
and the presence of the weak/strong noise component in the power spectra. 

GX 339-4 is a well-studied transient stellar-mass black-hole binary having a mass function of 
$M$ = $5.8\pm0.5~M_\odot$ (Hynes et al. 2003) and a low-mass companion of mass 
$0.52~M_\odot $ and being located at a distance $d~\geq~6$~kpc (Hynes et al. 2003, 2004). 
The inclination angle ($i$) of the system has not yet been confirmed. Zdziarski et al. (2004) 
indicated the lower limit of $i~\geq45^\circ$, while Cowley et al. (2002) indicated the upper limit of 
$i~\leq60^\circ$, although there is another report of an inclination angle of $i~\sim~20^\circ$ 
(Miller et al. 2004) of this source.

GX 339-4 has undergone several outburst phases (Nowak et al. 1999; Belloni et al. 2005; Motta et al. 2009, 
Debnath et al. 2010) during the {\it Rossi X-ray Timing Explorer} (RXTE) era (in the period from 1998 to 2011). 
Since its discovery (Markert et al. 1973), the source has been found to be in low luminosity (quiescent state) 
states, although there have been several reports of state transitions (Maejima et al. 1984; Ilovaisky et al. 1986).
As for other transient black-hole candidates (e.g., GRO J1655-40, XTE J1550-564, H 1743-322), 
the complex outburst profile of the source, generally, begins and ends in the low-hard state 
(Belloni et al. 2005; Debnath et al. 2008). 

This general behavior of the outburst profile can be explained using a two-component 
advective flow model (Chakrabarti \& Titarchuk 1995, hereafter CT95) having a standard 
Shakura-Sunyaev Keplerian disk (Shakura \& Sunyaev 1973) on the equatorial plane and 
a sub-Keplerian (low angular momentum) accreting halo on top of the Keplerian disk. 
The sub-Keplerian halo in the shocked-accretion phase produces the Compton cloud 
(inner part of the disk) of hot electrons, while the Keplerian disk produces the soft 
photons. The high energy part of the spectrum is due to the inverse Comptonization of 
the soft photons by the hot electrons supplied by the sub-Keplerian disk. In this scenario, 
the outburst phenomenon could be due to a sudden increase in the viscosity of the accretion-disk 
system (e.g., Frank, King \& Raine 2002). More specifically, Mandal \& Chakrabarti (2010) 
suggests that the enhanced viscosity converts part of the sub-Keplerian matter 
(i.e., low angular momentum flow) into a viscous Keplerian flow (i.e., matter in Keplerian 
rotation), keeping the total mass-flow rate roughly constant. The sudden increase in the 
accretion rate of the Keplerian matter and its inward movement cause the sudden enhancement 
of luminosity. The declining phase starts when the source of the enhanced viscosity is removed 
and there is a shortfall of the Keplerian component.

The typical evolution of the QPOs during the outburst phases of the transient black-hole sources 
has been well-established for a long time (Belloni \& Hasinger 1990; Belloni et al. 2002; 
Belloni et al. 2005; Debnath et. al 2008, 2010, 2012; Chakrabarti et al. 2005, 2008, 2009).
The outbursting BHCs in general show signatures of low and intermediate frequency 
quasi-periodic oscillations (QPOs) during the initial rising phase (mainly low-hard state) 
as well as in the decline phase of the outburst, where the sub-Keplerian rate dominates and during 
the intermediate state, where the Keplerian and sub-Keplerian rates are more or less comparable to 
each other. In general, during the rising phase (hard state) of the outburst, the frequency of 
the QPO increases, whereas during the declining phase, the QPO frequency gradually decreases. 
These QPO types of evolution (increasing/decreasing nature of QPO frequencies) in these 
objects can be well-understood due to the propagating oscillatory shocks 
(POS; Chakrabarti et al. 2008, 2009, Debnath et al. 2010) model. 

After remaining in the `quiescent' state for almost three years (the last outburst took place 
during the year 2006/2007, although there is report of weak activity in 2009 as observed 
by SWIFT/BAT), GX 339-4 showed X-ray flux activity of 17 mCrab (4-10 keV) on January 03, 
2010 with the first detection by MAXI/GSC onboard ISS (Yamaoka et al. 2010). The source remained 
active in X-rays for the next $\sim 430$ days and during this period, the source was extensively 
monitored with RXTE, starting from 2010 January 12, (Tomsick, 2010). In Debnath et al. 
(2010) (hereafter, Paper I), we provided a preliminary summary of the timing and spectral 
properties during the rising phase of this outburst.

In this present paper, we provide the timing and spectral results of RXTE PCA for a total of 
$236$ observational intervals of $419$ days of data, from $2010$ January $12$ to $2011$ March $6$. 
From our study, a comprehensive understanding of the mass accretion processes and 
properties of the accretion disk of this black hole candidate
has emerged. The PCA spectra of $2.5-25$ keV were mainly fitted with the combination of two 
components of a disk black body and a power law whereas the soft state spectra were fitted across 
the energy range of $2.5-10$ keV (few spectra extended up to $15$ keV also) because of very low 
and insignificant flux contributions above $10$ keV. On the basis of the relative importance of 
the black body and the power-law components, the outburst is divided into four spectral 
states, namely, {\it hard, hard-intermediate, soft-intermediate}, and {\it soft} (Homan \& Belloni 2005),
in the sequence of {\it hard $\rightarrow$ hard-intermediate $\rightarrow$ soft-intermediate 
$\rightarrow$ soft $\rightarrow$ soft-intermediate $\rightarrow$ hard-intermediate $\rightarrow$ hard}.
Since the definitions vary, we here define the hard state to be the one in which the temperature
($T_{in}$) and photon index ($\Gamma$) parameters have values greater than $\sim 1.5$~keV and 
less than $\sim 1.6$ respectively, whereas in the hard-intermediate state (i.e., in rising and 
declining phases) the parameters vary rapidly within the ranges of $\sim 1.5-1.0$~keV and $\sim 1.6-2.3$.
In the soft-intermediate state, the $T_{in}$ and $\Gamma$ parameters have values within the range of 
$\sim 0.9-0.7$~keV and $\sim 2.3-2.5$, whereas in the soft state the values are $< 0.7$~keV and $\sim 3.0$, 
respectively. Debnath et al. (2012) also reported the similar and cyclic order of state transitions in 
the 2010 and 2011 outbursts of the transient Galactic black-hole candidate H~1743-322. 
The temporal variations in these states and state transitions are also found to be unique 
and tightly correlated with the spectral properties.  

The paper is organized in the following way: In the next section, we discuss the observations
and data analysis procedures using HEASoft software. In Sect. 3, we present the details of our
temporal and spectral analysis of PCA data and discuss possible accretion-disk flow
behavior during the outburst phase. Finally, in Section 4, we present a brief discussion 
of our results and provide some concluding remarks.

\section {Observation and data analysis}

We present our analysis of publicly available archival data from the RXTE 
Proportional Counter Array (PCA) instrument of the entire 2010-11 outburst of GX 339-4. 
We extracted and analyzed the RXTE archival data from $2010$ January $12$ (modified Julian date 
(MJD) $55208$) to $2011$ March $6$ (MJD $55626$) from the PCA (Jahoda et al., 1996). 
We extracted light curves and PDS for different energy bands and the energy spectra from the good 
and the most reliably calibrated detector units {\it i.e.}, PCU2, for the PCA. We also analyzed 
RXTE/ASM data to have a quick look at the outburst profile. The results of our analysis using the 
ASM data were published in Debnath et al. (2010). Here, we only present results based on the analysis 
of the archival data of the PCA instrument.  

We carry out our data analysis using the FTOOLS software package HeaSoft version HEADAS-6.10 
and XSPEC version 12.6. For the timing analysis, we use the science data of the {\it Event mode} 
($E\_125us\_64M\_0\_1s$, FS37*.gz) with a maximum time resolution of $125~\mu s$.
We use standard FTOOLS tasks (`xtefilt', `maketime') to generate both a filter file using the 
latest pca breakdown history file and a good time interval (gti) file for PCU2. To determine the 
`good time interval' for each observation, we use the screening criteria of elevation angle $> 10^\circ$, 
offset $< 0.02^\circ$, SAA passage time (within 30 min of passage), and HV breakdown time for the PCU2 
detector only. Light curves were extracted using the ``sefilter" and ``saextrct'' task for the event and 
science data, respectively. 

For spectral analysis, {\it Standard2} mode Science Data of PCA (FS4a*.gz) are used. Spectra are
extracted from all the layers of the PCU2 for 128 channels (without any binning/grouping the channels). 
We exclude the HEXTE data from our analysis, 
as we find strong residuals (line features) in the HEXTE spectra at different energies. This  is 
possibly due to the difficulties in estimating the background spectra as the `rocking' mechanism has 
stopped functioning for HEXTE. So, we restrict our spectral analysis for the PCA data for energy range 
of $2.5 - 25$ keV, although we analyzed the PCA spectral data in a few observations (for energies 
up to $40$ keV) to search for the high energy contribution to the low-hard states.
The ``runpcabackest" task was used to estimate the PCA background using the latest bright-source 
background model (count rates were always high, except for the last few observations of the decline 
phase, where the net count rates per PCU were $\leq 40$ counts/sec). We also incorporated the 
$pca\_saa\_history$ file to take care of the SAA data. To generate the response files, we used 
the ``pcarsp" task. In general, we follow the same analysis techniques discussed in Paper I 
(Debnath et al. 2010) for the timing and spectral analysis purposes.

\section {Results and modeling}

The study of the X-ray temporal and spectral properties of compact objects, specially for outbursting 
BHCs, is essential to determine the flow properties during the outburst phases of the sources. It was 
pointed out by Debnath et al. (2010) that there are two main types of outbursting BHCs, one of so-called 
type fast-rise slow-decay (FRSD) and another of slow-rise slow-decay (SRSD) (i.e., GX 339-4), although 
the general behavior of outbursting X-ray binaries is far more complex. Chen et al. (1997) carried out 
an extensive study of the light curve morphology of several X-ray binaries and classified them into five 
morphological types. To study the X-ray intensity variations of the 2010 outburst of GX 339-4, we extracted 
light curves from PCU2 data acquired by the RXTE/PCA instrument in the different energy bands $2-6$ keV 
($5-13$ ch.), $6-20$ keV ($14-46$ ch.), and $2-20$ keV ($5-46$ ch.). We divided the $2-20$ keV energy band 
into the above two bands because the $2-6$ keV photons mainly come from a thermally cool Keplerian disk, 
whereas the photons in the higher energy band ($6-20$ keV) band come from a Comptonized sub-Keplerian disk 
(Compton corona). 

In Figs. 1(a-b), the total $2-20$ keV PCU2 light curve (counts/sec) and the hardness ratio (ratio 
of the photon count rates in the $6-20$ keV to $2-6$ keV bands) are plotted. The origin of the 
time axis is MJD 53200 ($2010$ January $4$), which is eight days prior to our first observation. 
The hardness ratio variation distinctly reflects the indication of spectral state transitions. 
The hard to hard-intermediate state transition took place on $2010$ April $10$ (MJD 55296), 
the hard-intermediate to soft-intermediate transition on $2010$ April $18$ (MJD 55304), 
the soft-intermediate to soft transition on $2010$ May $15$ (MJD 55331), the reverse transition 
from the soft to soft-intermediate state on $2011$ January $4$ (MJD $55565$), the soft-intermediate 
to hard-intermediate state transition on $2011$ February $2$ (MJD $55594$), and finally on 
$2011$ February $14$ (MJD $55606$) the state transition from hard-intermediate to hard state occurred. 
These are indicated by vertical dotted lines in the plot of Fig. 1. It is observed that rapid changes 
in the hardness ratio occur only in the hard and hard-intermediate states, whereas in the 
soft and soft-intermediate states the hardness ratio changes only very slowly. Both the rising and 
falling arms of the outburst profile correspond to the hard state and in both the cases, we found 
evidence of strong QPOs. However, the local changes in the temporal and spectral features in different 
states cannot be discerned from this plot. This leads us to conduct a thorough and robust temporal and 
spectral analysis using the PCA data and the results are presented in the following sections.

The source was in a hard state from our initial RXTE/PCA observation day ($2010$ January $12$, MJD $55208$) 
to $2010$ April $9$ (MJD $55295$). During this initial rising phase, no QPOs were detected until 
$2010$ March $21$. On 2010 March $22$, for the first time a QPO of $102$~mHz frequency was observed. 
After that, it increased monotonically until $2010$ April $17$, when we observed a QPO of 
$5.692$~Hz. The source moved to a hard-intermediate spectral state on $2010$ April $10$ (MJD $55296$), 
where it remained until $2010$ April $17$ (MJD $55303$). On $2010$ April $18$ (MJD $55304$), the source 
moved to a soft-intermediate spectral state and remained in this state until $2010$ May $14$ (MJD $55330$). 
During this spectral state, QPOs were observed sporadically on and off at $\sim 6$~Hz. After that on 
$2010$ May $15$ (MJD $55331$), the source moved to a soft spectral state, during which no QPO was observed. 
It remained in this state for around next seven and a half months, until $2011$ January $3$ (MJD $55564$). 
During this time, it emitted a hard X-ray flux ($> 10$ keV) dominated by a sub-Keplerian flow 
component that diminished rapidly with time to produce spectral data of too low signal-to-noise ratio 
above $\sim 15$ keV. On $2011$ January $4$ (MJD $55565$), the source underwent a spectral transition 
to the soft-intermediate state as the flux became increasingly hard. During this spectral state, 
sporadic QPOs of frequency $\sim 2$~Hz were observed, as in the same state of the initial rising 
phase. After this time, the source returned to a hard-intermediate state on $2011$ February $2$ 
(MJD $55594$). During this phase, QPOs of monotonically decreasing frequency were observed, as for 
other transient BHCs (e.g., the 2005 outburst of transient Galactic BHC GRO J1655-40; Chakrabarti 
et al. 2008, Debnath et al. 2008 and the 
2010 and 2011 outbursts of H 1743-322; Debnath et al. 2012). After that, the source returned to its 
hard state on $2011$ February $14$ (MJD $55606$). Here we should also observe QPO evolutions. However,
owing to decreases in their photon fluxes and a lack of long duration observations, low-frequency 
QPOs have not been detected prominently, only break frequencies being observed (see, Table 1).

\begin{figure}
\vbox{
\vskip -0.0cm
\centerline{
\includegraphics[scale=0.6,angle=0,width=8truecm]{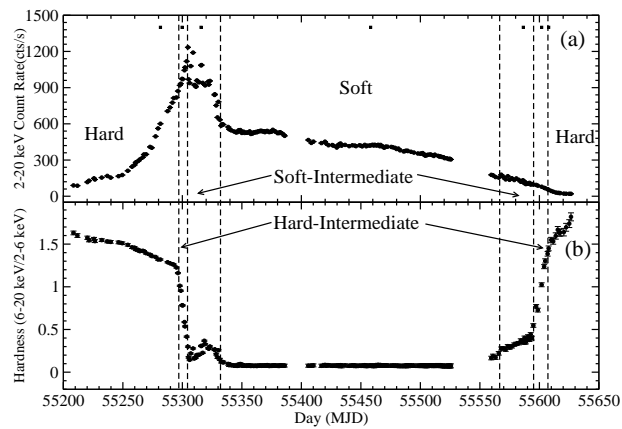}}
\vspace{0.0cm}
\caption{(a) 2-20 keV PCA light curve and (b) hardness ratio (6-20 keV versus 2-6 keV count ratio) 
as a function of the MJD of the event (i.e., 2010-11 outburst of GX 339-4).
The vertical dashed lines indicate the transition of states and filled squared points in (a) 
are the seven observed days from seven different spectral states, whose detailed timing and 
spectral results are discussed in this paper.}}
\label{kn : fig1}
\end{figure}

To clearly monitor the evolution of states during the outburst, we need to draw the HID and study 
the evolution of timing and spectral properties during different branches of the HID.
In Fig. 2, we plot the PCA $2-20$ keV count rate of the 2010-11 outburst against X-ray color 
(PCA count ratio of the flux in the $6-20$ keV to that in the $2-6$ keV energy band). The points 
$A$ and $H$ are the indicators of the start and the end of the RXTE observations respectively, 
whereas the points $B$, $C$, $D$, $E$, $F$, and $G$ are the points on the days where the state 
transitions occurred. 

\begin{figure}
\vbox{
\vskip -0.2cm
\centerline{
\includegraphics[scale=0.6,angle=0,width=8truecm]{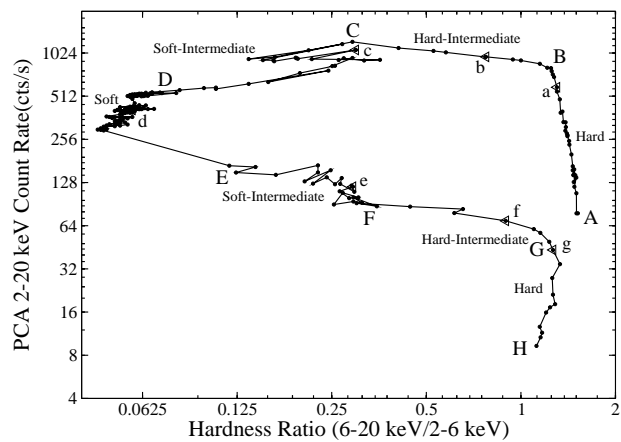}}
\vspace{0.0cm}
\caption{Hardness intensity diagram of GX 339-4 in the 2010-11 outburst observed 
with RXTE/PCA. We have plotted the total count rates in $2-20$ keV energy band along Y-axis 
and the ratio of the count rates in the $6-20$ keV to $2-6$ keV bands along the X-axis. 
The points $A$, $B$, $C$, $D$, $E$, $F$, $G$, and $H$ are observations on MJD $55208$, 
MJD $55296$, MJD $55304$, MJD $55331$, MJD $55565$, MJD $55594$, MJD $55606$, and 
MJD $55626$, respectively, which indicate the start/state transitions/end of our observation. 
In addition, the points $a$, $b$, $c$, $d$, $e$, $f$, and $g$ (marked with triangle on plot) 
are the seven observed points from seven different states, whose detailed timing and spectral 
results are discussed in this paper. Note that the X and Y axes are on logarithmic scales.}}
\label{kn : fig2}
\end{figure}

In Figs. 3(a-g), we plotted the light curves and color-color diagrams for seven data sets (a-g, 
marked in Fig. 2) to study the X-ray variability and hardness ratio variation in different branches 
of the HID. We extracted PCA light curves for the above seven data sets using time bins of 1 sec 
in three different energy channels: $I$ consisting of $5-8$ channels ($2-4$ keV), $II$ of $9-35$ 
channels ($4-15$ keV), and $III$ of $36-58$ channels ($15-25$ keV). Using these light curves, we 
generated a total light curve for the $2-25$~keV energy range and two X-ray colors defined as 
HR2 ($III/I$) versus HR1 ($II/I$). Light curves were then background subtracted. The spectral data 
shows that the background rates are $\sim$ 10 to 15 cts/sec in the energy range of $2-25$ keV. 
Our motivation for splitting the energies in to the above mentioned intervals was to separate the 
contributions of photons coming from the Keplerian and sub-Keplerian disk components. 
The photons from the Keplerian disk are emitted primarily at low energies ($< \sim4$ keV),  
whereas the same from the sub-Keplerian flow were emitted at higher energies ($> \sim 4$ keV) 
for the stellar-mass black-hole candidates. Thus, $I$ is emitted mostly from the Keplerian 
component, whereas the component $II$ might be emitted from the region where the moderate 
thermal Comptonization of the Keplerian photons take place. The component $III$ should be 
emitted from the region that is definitely depleted or enhanced during state transitions 
as it represents the higher energy side of the pivotal energy [$\sim 15$ keV] in the spectrum. 
Thus, these figures carry some information about the disk geometry, i.e., the number of 
soft photons produced by the Keplerian disk ($\sim I$) and the number of Comptonized 
photons that produced by the `Compton cloud' [$\sim (II + III)$]. 

\begin{figure}[h]
\vbox{
\vskip -0.0cm
\centerline{
\includegraphics[width=3.6in,angle=0]{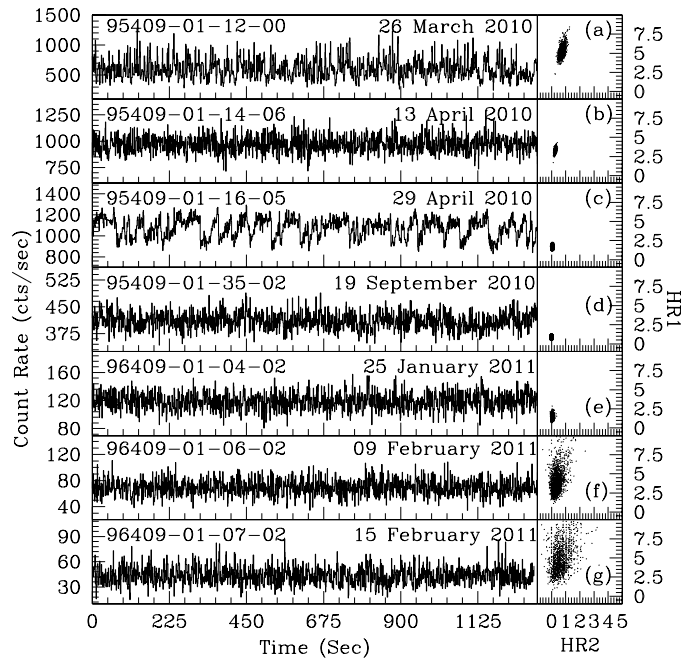}}\vskip 0.1cm
\caption{(a-g): In each plot, we plot in the left panel the 2-25 keV PCA light curve 
and in the right panel the color-color diagram are plotted. These plots (a-g) are 
associated with MJD 55281, MJD 55299, MJD 55315, MJD 55458, MJD 55586, 
MJD 55601, and MJD 55607, respectively and represent seven different spectral 
states of the outburst of GX 339-4.}}
\label{kn : fig3(a-f)}
\end{figure}

Figure 3(a-d) represent the hard, hard-intermediate, soft-intermediate, and soft states, 
respectively, of the rising phase of the outburst. The corresponding color-color diagrams 
show that the photon intensity of the harder component ($III$) is strongest in the hard state 
and slowly decreases as the outburst moves to a soft state. However, the soft component ($I$) 
has the opposite behavior, namely it brightens and emits its strongest flux in the soft state.
The photon intensities of bands $III$ and $I$ are comparable to each other the in hard-intermediate 
state. The power density spectrum of this hard state data shows a strong signature of a QPO 
at $0.134$ Hz, whereas that of the hard-intermediate state shows a QPO at $2.423$ Hz with harmonics 
at $4.853$ Hz and $7.249$ Hz.

A B-type QPO of frequency $\sim 6$~Hz is observed in the soft-intermediate state. Interestingly, 
we observe variability of longer timescale (few tens of seconds) for some obs-IDs in the 
soft-intermediate state. A similar kind of variability was also observed in the very high state (VHS) 
of GX 339-4 with timescales of $20$ to $60$ sec (Miyamoto et al. 1991). The dynamic nature of this type 
of variability will be studied in detail and presented elsewhere. In the soft state, the sub-Keplerian 
disk becomes cooler, a greater supply of Keplerian matter, and we observe no signatures of QPOs.

Figure 3(e-g) again shows the soft-intermediate, hard-intermediate, and hard spectral state, 
respectively, but in the declining phase of the outburst. From the color-color variations, we 
note that the soft-photon intensity ($I$) decreases and the hard component of the flow ($III$) 
becomes stronger. The soft-intermediate data shows a QPO of frequency $2.136$~Hz. 
The hard-intermediate and hard states display similar variations to the same states in the 
rising phase of the outburst, except that the X-ray intensities (counts/sec) in different energy 
bands are somewhat weaker. A strong QPO of $1.322$~Hz was observed in the hard-intermediate state, 
whereas in the hard state a weak QPO of break frequency type was observed (see Table 1 for details).

In the subsequent subsections, we present the modeling of the power spectral evolution, 
the QPO characteristics during the outburst phase, and the spectral energy distribution. The detailed
temporal and spectral analysis results of the entire observation are tabulated in online tables 
(A.1, A.2, and A.3).

\subsection {Power density spectra}

In Sect. 3, we have examined the PCA light curves and color variations in different branches of 
the HID. This motivated us to carry out a similar type of examination in power spectra in each data 
set to obtain a clearer picture of the disk dynamics following the HID variations. Power spectra were 
produced with a time resolution of $1/2048$ sec (using the event mode data of $122 \mu$s time resolution), 
which corresponds to a Nyquist frequency of $1024$ Hz. For each observation, we fit the average spectra 
(energy range of $2-15$ keV) with a constant to estimate the white noise component (Poissonian noise). 
We found that the expected white noise level (above $30$ Hz) in hard states has a power of $\sim 1.97$, 
whereas in soft states the level was around $2$. During the PDS analysis, we did not include the dead time 
effects because the count rates are not so high (maximum rate around $1200$ cts/sec). We verified that the 
error caused by the PCA dead time is at most 4\%. We used the ``powspec" task of the XRONOS package with 
a normalization factor of `-2' to ensure that the expected `white' noise was subtracted from the 
rms fractional variability of PDS. The power has the unit of rms$^2$/Hz. 
We re-binned the power spectrum with a geometrical factor of $-1.05$ to have nearly equispaced 
logarithmic frequency bins and fitted each spectrum within the range from $0.01$ to $20$ Hz.

\begin{figure}[h]
\vbox{
\vskip -0.2cm
\centerline{
%\hskip -0.5cm
%\includegraphics[height=2.6in,width=2.4in,angle=0]{Figures/4_State_PDS.ps}}
\includegraphics[height=2.6in,angle=0]{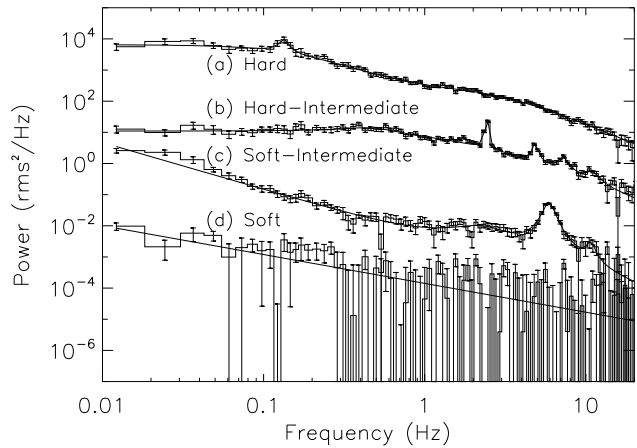}}
\caption{Model-fitted PDS of four spectral states (a, b, c, and d in Fig. 2) are shown. From top to bottom: 
hard state data of $2010$ March $26$ (Obs. ID: 95409-01-12-00), hard-intermediate state of $2010$ April 
$13$ (Obs. ID: 95409-01-14-06), soft-intermediate state of $2010$ April $29$ (Obs. ID: 95409-01-16-05), 
and soft state of $2010$ September $19$ (Obs. ID: 95409-01-35-02) PDS are shown. For a clear view, the Y-scale of 
(a), (b), and (c) plots are multiplied by the factor of $2 \times 10^4$, $10^3$, and $20$, respectively.}}
\label{kn : fig4(a-b)}
\end{figure}

In Fig. 4(a-d), we present power spectra from four different regions (labeled a, b, c, and d) 
of Fig. 2. The power spectra are generally dominated by flat-top (white noise) and power-law like 
(red noise) components, which have typically been fitted with a zero-centered Lorentzian 
(Belloni, Psaltis \& van der Klis 2002b). The presence of QPOs in the power spectra has 
also been fitted by a Lorentzian (Nowak 2000, van der Klis 2005). We fitted all the PDSs with 
either Lorentzian profiles or power-law like distributions in the frequency range from 
$0.01$ to $20$ Hz. The overall fitted results of the power spectra along with the best-fit 
parameter values for the QPOs in all four cases are given in Table 1. After fitting PDS, 
we used the ``fit err" task to calculate the error in the QPO frequencies, widths, 
and powers. This task calculates the 90\% confidence range of any fitted parameter. 

In Fig. 4(a), we present the Lorentzian-model-fitted power spectrum of the light curve of 
$2010$ March $26$ (Obs. ID: 95409-01-12-00), which belongs to the {\it hard} state of the HID 
(the $A$ to $B$ branch) and also contains the prominent signature of a type-C QPO. In this
branch of the HID, the power spectra are generally fitted with three Lorentzians - (i) in the 
flat-top noise part plus a power-law distribution up to the break frequency (with a zero-center 
Lorentzian); (ii) at QPO frequencies; and (iii) for the high-frequency power law after the 
break frequency. The break frequency and QPOs observed in the power spectra vary, respectively, 
from $0.5$ to $2.0$ Hz and from $0.102$ Hz to $0.547$ Hz as the source moves from hard to 
hard-intermediate states. All observations in this state are those of a strong `C' type QPO.

The model-fitted power spectra in Fig. 4(b) belong to the {\it hard-intermediate} state and were 
observed on $2010$ April $13$ (Obs. ID: 95409-01-14-06). A strong flat-top noise component
along with a characteristic break frequency at $0.44$ Hz and QPO frequency around $2.42$ Hz with 
harmonics are discernible in the PDSs, which in this branch of the HID ($B$ to $C$) are also fitted 
with the same Lorentzian components as in the previous observations. The QPOs observed in this branch
are of type-C and the QPO frequency increased as the source moved towards the soft-intermediate state.
We observe that the flat-top noise component diminished and the power-law like distribution started to 
dominate the power density spectra in the low frequency range as the source moved towards the 
soft-intermediate state.

In Fig. 4(c), we present the model-fitted PDS for observation performed on $2010$ April $29$ 
(Obs. ID: 95409-01-16-05), which corresponds to a {\it soft-intermediate} state. 
A strong QPO of type-B is observed in the PDS. The PDSs are generally fitted with power-law 
and Lorentzian-type profiles. In this branch of the HID (from $C$ to $D$), the QPOs are observed 
sporadically at around $6$ Hz.

In Fig. 4(d), we present the power spectra of the observation on $2010$ September $19$ 
(Obs. ID: 95409-01-35-02), which belongs to the {\it soft} state of the HID ($D$ to $E$). 
In the soft states, no QPOs were observed and the power spectra were mostly dominated by 
a single power-law like distribution of slope $\sim~-1$.

The power spectra in the bottom branch of the HID ($E$ to $F$, $F$ to $G$, and $G$ to $H$) belong to 
the soft-intermediate, hard-intermediate, and hard state of the decline phase, which have the same 
features as previous observations in the top branches ($D$ to $C$, $B$ to $C$, and $A$ to $B$) of the 
HID, but with less variability as the photon count rate (see, Fig. 2) has decreased significantly. The 
model-fitted parameters and calculated {\it rms} power of all of these three states are given in Table 1.

\begin{table}
\vskip -0.2cm
\addtolength{\tabcolsep}{-4.00pt}
\scriptsize
\centering
\centerline {Table 1}
\centerline {Temporal properties of power spectra in different states}
\begin{tabular}{lcccccccc}
\hline
\hline
Obs.& $\nu$&$\Delta\nu$&Q&Type&$\nu_{brk}$&Noise/Comp.& rms($\%$) &$\chi^2$/DOF \\
    (1) & (2) &  (3)  &  (4)  &  (5)  &  (6) & (7) & (8) & (9) \\
\hline
26 &$0.134_{-0.006}^{+0.006}$&$0.022_{-0.002}^{+0.001}$&6.091&C&0.88& lo + lo + lo & 36\% & 77.76/81 \\
\hline
\multirow{3}{*}{35} &$2.423_{-0.006}^{+0.006}$&$0.072_{-0.007}^{+0.006}$&33.65&C&0.44& & &\\
 &  $4.853_{-0.036}^{+0.029}$&$0.412_{-0.120}^{+0.127}$&11.78&C&...& lo+lo+lo+lo+lo& 19\% & 108.64/76\\
 &  $7.249_{-0.137}^{+0.122}$&$1.256_{-0.446}^{+0.724}$&5.771&C&...& & &\\
\hline
51 &$5.908_{-0.037}^{+0.038}$&$1.021_{-0.083}^{+0.086}$&5.786&B&2.122& po+lo+lo+lo& 11\%  & 113.9/79\\
\hline
141 &...&...&...&...&...& po & 3\% & 96.39/88\\
\hline
208 &$2.136_{-0.160}^{+0.138}$&$1.065_{-0.068}^{+0.087}$&2.005&B&...& po+lo&11\% & 77.17/85\\
\hline
222 &$1.322_{-0.024}^{+0.002}$&$0.162_{-0.056}^{+0.053}$&8.160&C&0.331& lo+lo+lo & 23\% & 73.73/81 \\
%222&$1.322_{-0.024}^{+0.002}$&$0.163_{-0.056}^{+0.053}$&8.080&C&0.331& lo+lo+lo & 5\% & 73.73/81 \\
%\multirow{3}{*}{222}&$1.306_{-0.011}^{+0.004}$&$0.354_{-0.021}^{+0.006}$&3.689&C&0.331& lo+lo+lo & 5\% & 73.73/81 \\
% &  $2.906_{-0.371}^{+0.275}$&$2.152_{-0.375}^{+0.214}$&1.350&C&...& & & \\
\hline
226 & ... & ... & ... &...&0.133& lo+lo &24\% &92.14/84 \\
\hline
\end{tabular}

\leftline {The {\it rms} is calculated for each power spectra from $0.01$ to $20$ Hz.}

\end{table}

The evolution of broad aspects of the power spectra in different states could be well-understood 
by studying the diffusive propagation of the perturbation in the disk-like configuration (by treating 
a Keplerian disk as an extended disk and a sub-Keplerian Compton cloud as an inner disk) 
(Shakura \& Suyanev 1973; CT95; Titarchuk, Shaposhnikov \& Arefiev 2007, hereafter TSA07). 
TSA07 showed that the entire power spectra can be modeled with a `white-red-noise' (WRN) continuum, 
which has two components - a low-frequency (LF) power-law like distribution and a high-frequency (HF)
flat-top (white noise) along with a red-noise (power-law type) component. The HF-WRN continuum is 
characteristic of the hard state, whereas the LF-WRN continuum represents the soft state. 
In Fig. 4, the power spectra of the hard and hard-intermediate states are well-described by 
the HF-WRN continuum and the soft state spectra are only modelled by LF WRN (power-law) continuum. 
The soft-intermediate power spectra are composed of both LF and HF-WRN components. The detailed
modeling of the power spectra of this source is beyond the scope of the present work. 

\subsection {QPO evolution and the POS model-fitted results}

The evolution of the low and intermediate QPOs in the outburst sources has been reported quite 
extensively in the literature (Belloni \& Hasinger 1990; Belloni et al. 2002; Belloni et al. 2005; 
Debnath et. al 2008, 2010, 2012; Chakrabarti et al. 2005, 2008, 2009). It has been shown for several 
outburst sources (namely BHs) that the evolution of QPOs can be well-explained by the theoretical 
propagating oscillatory shock (POS) model (for details, see Chakrabarti et al. 2008, 2009; 
Debnath et al. 2010, 2012). 

We found that our detailed PDS analysis detected QPOs in a total of $28$ observations performed over 
in $27$ days during the rising phase and in a total of $22$ observations performed over $22$ days 
during the declining phase of the outburst. Figure 5 shows the variation in the day-wise QPO 
frequencies i.e. QPO evolutions during (a) the initial rising phase and (b) the final declining 
phase of the recent 
GX 339-4 outburst. In both cases, these QPO evolutions are fitted with a POS model. According to this 
model (POS; Chakrabarti et al. 2008, 2009; Debnath et al. 2010, 2012), we can derive from the observed 
QPO frequency ($\nu_{QPO}$) an idea about the location of the shock wave ($r_s$), as these shock 
oscillations are believed to be responsible for the generation of the QPOs. The QPO frequency 
($\nu_{QPO}$) is proportional to the inverse of the light crossing time ($t_{infall}$) from the shock 
location to the black-hole, i.e., $\nu_{QPO}~\sim~(t_{infall})^{-1}$ and also 
$t_{infall}~\sim~R~r_s~(r_s - 1)^{1/2}~\sim~r_s^{3/2}$, where $R$ is the shock compression ratio 
(= $\rho_+$/$\rho_-$, where $\rho_+$ and $\rho_-$ are the densities in the post- and the pre- shock 
flows). The QPO frequency in this model is $\nu_{QPO}~\sim~r_s^{-3/2}$. In a propagating shock 
scenario, $r_s=r_s(t)$, is the time-dependent shock location given by $r_s(t)=r_{s0} \pm v t/r_g$, 
where $v$ is the velocity of the shock wave, the `-' sign indicates the rising phase, the `+' sign 
represents the declining phase of the outburst, and $r_s$ is measured in units of the Schwarzschild 
radius $r_g= 2GM/c^2$ ($G$ is the gravitational constant, $M$ is the BH mass, and $c$ is the velocity 
of light). For the constant movement of the shock wave (as in the rising phase of the recent GX 339-4 
outburst) $v=v_0$ and for the accelerating case (as in the declining phase of the recent outburst) 
$v=v(t)$, can be defined as $v(t)=v_0 + a t$, where $a$ is the 
acceleration of the shock front.
 
\begin{figure}[h]
\vbox{
\vskip -0.1cm
\centerline{
\includegraphics[scale=0.6,angle=0,width=07truecm]{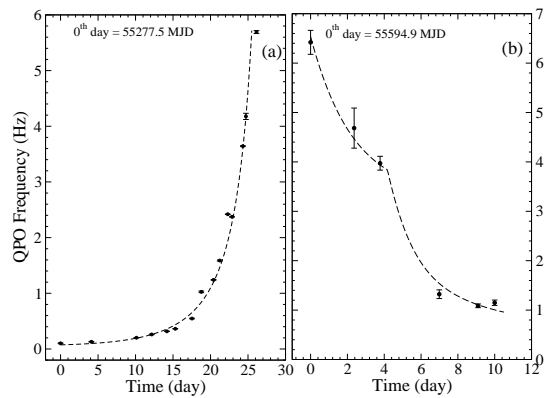}}
\caption{ Variation of the QPO frequencies with time (in day) during (a) the rising and (b) the 
declining phases of the outburst. The dotted curves are the solutions from the oscillating and 
propagating shocks. While in Fig. (a), the shock appears to be drifting at a constant speed 
($\sim 10~m~s^{-1}$) towards the black hole, in Fig. (b), the shock moves away from the black 
hole in two different ways. During the initial $\sim 4.2$~days, the shock moved away with a 
slow rate of acceleration ($\sim 20~cm~sec^{-1}~day^{-1}$), after which it moved away with a 
high acceleration ($\sim 175~cm~sec^{-1}~day^{-1}$).}}
\label{kn : fig5}
\end{figure}

Figure 6 shows the change in the shock locations during (a) the rising and (b) the declining 
phases of the outburst. In Fig. 6(a), the variation in the compression ratio $R$ (the inverse 
of the shock strength $\beta$) is also shown, which varies from $4$ (the strongest possible shock) 
to $\sim 1$ (the weakest possible shock) during the QPO evolving period of $\sim 26$ days. Similarly 
during the declining phase of the outburst, $R$ started from the strongest possible shock 
value (=$4$), after which it became weaker as the day progressed, and reaching its lowest possible 
value ($R \sim 1$) on about the tenth day ($2010$ February $12$, MJD $55604$), when a $1.149$~Hz QPO 
was observed.  

The details about the rising-phase QPO evolution are given in {\it paper I} (Debnath et al. 2010), 
so we present a brief report here. During this phase, the shock moves towards the black hole with 
a constant speed of $\sim 10~m~s^{-1}$ from $1500~r_g$ on the first day that the QPO is observed 
($2010$ March $22$, MJD 55277.5), where a $102$~mHz frequency QPO was indeed observed until 
$172~r_g$ on $2010$ April $17$ (MJD 55303.6), when a $5.692$~Hz QPO was observed with a period 
of $\sim 26$ days. During this QPO evolutionary phase, the source remains in either a hard or 
hard-intermediate spectral state. After that, the source moved to a soft-intermediate 
spectral state and QPOs were observed sporadically at $\sim 6$~Hz, remaining in this state 
for the next $\sim 26$ days. It is found that the observed QPOs are of type-C in the hard 
and hard-intermediate states, whereas they are of type-B during the soft-intermediate spectral state. 

During the declining phase of the outburst, the source was observed in the soft-intermediate state 
(from $2011$ January $4$, MJD $55565$ to $2011$ February $1$, MJD $55593$) just before the hard state, 
when QPOs were observed at $\sim 2$~Hz. After that, during the hard-intermediate state (from $2011$ 
February $2$, MJD $55594$ onwards), the QPO frequencies monotonically decrease from $6.42$~Hz 
(on $2011$ February $2$, MJD $55594$) to $1.149$~Hz (on $2011$ February $12$, MJD $55604$). 
After that, during the hard state no prominent QPOs were observed owing to the decrease in source 
flux and lack of long-duration observations, only break frequency type QPOs being observed. 
In our POS model solution, we break the present QPO evolution period of hard-intermediate state into 
two parts. During the initial $4.2$ days, the shock moved away from the black hole from $84~r_g$ to 
$155~r_g$ with a slow variation in shock velocity (from $205~cm~sec~^{-1}$ to $288~cm~sec^{-1}$) 
and slow rate of acceleration ($20~cm~sec^{-1}~day^{-1}$). After that, the shock wave receded 
at a higher rate of acceleration ($175~cm~sec^{-1}~day^{-1}$), reaching $751~r_g$ on the last day 
($\sim 10$th day) of our POS model fit, where a $1.149$~Hz QPO was observed. On the last day of the 
QPO evolution, the shock velocity was found to be $\sim 1785~cm~sec^{-1}$. As in the rising phase, 
during the declining phase of the QPO evolution, we started with a strong shock ($\beta = 0.25$, 
i.e., $R=4$), which became weaker with time reaching its minimum possible value of $R=1$. The 
compression ratio $R$ decreased with time according to the relation 
$1/R \rightarrow 1/R_0 + \alpha (t_d)^2$, where $R_0$ is the initial compression ratio (here $R_0=4$), $t_d$ is the time in days (assuming that the 
first observation day is the zeroth day), and $\alpha$ is a constant that determines how the shock 
(strength) becomes weaker with time. During the initial $4.2$ days of the declining phase of QPO evolution, 
$R$ decreased slowly from $4$ to $2.74$, while $\alpha$ = $0.0065$, and then on about the tenth day 
it reached $\sim 1$ for different $\alpha$ (=$0.02$).

\begin{figure}[h]
\vbox{
\vskip -0.1cm
\centerline{
\includegraphics[scale=0.6,angle=0,width=08truecm]{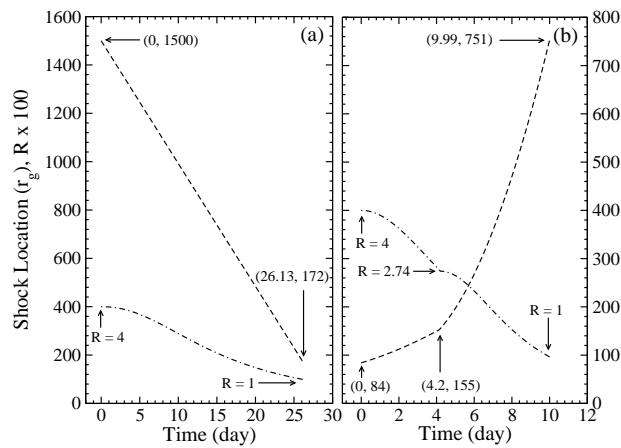}}
\caption{Variation in the shock locations (dashed curve) and compression ratios $R$ (dotted-dashed 
curve) with time (in days) during (a) the rising and (b) the declining phases of the outburst. See 
text, for details.}} 

%(???) In the rising phase, the shock moves towards the black hole from $1500~r_g$ to $172~r_g$ 
%within a period of $\sim 26$ days at a constant velocity, but with varying shock strength 
%($\beta \sim 1/R$). During the initial $4.2$ days of the declining QPO evolution phase, the shock 
%moves away from the black-hole from $84~r_g$ to $155~r_g$ with a slow acceleration 
%($20~cm~sec^{-1}~day^{-1}$), after which it moves away with higher acceleration 
%($175~cm~sec^{-1}~day^{-1}$) and reaches $\sim 751~r_g$ on the $\sim 10$th days. (???)}}
\label{kn : fig6}
\end{figure}

During both the rising and the declining phases of the outburst, a similar propagation of shock waves was 
observed during the rising and declining phases of the $2005$ GRO J1655-40 outburst (Chakrabarti et al. 2008). 
During the rising phase of the outburst, it was observed to have a monotonically increasing QPO frequency 
from $82$~mHz to $17.78$~Hz within a period of $\sim 16$ days with the constant movement 
($v_{s}~\sim~20~m~s^{-1}$) of the shock wave towards the black hole from $1270~r_g$ to $59~r_g$. 
During this QPO evolution phase, the shock compression ratio $R$ reached its weakest possible value 
($R~=~1$), after achieving its strong shock value ($R~=~4$). On the next day onwards, there was no 
detection of QPOs, hence it was concluded that there may be either no shock or a standing shock. However 
during the present GX 339-4 outburst in the soft-intermediate state, shocks might have stalled and 
oscillated randomly because QPOs were sporadically observed at $\sim 6$~Hz for the next $\sim 26$ days. 
This occurred when the rates of the two flows were comparable and the effects of the shock need not
always be visible. On the other hand, during the declining phase of the $2005$ GRO J1655-40 outburst, 
the shock moved away from the black hole starting from $40~r_g$ to $3100~r_g$ with a monotonically 
decreasing QPO frequency from $13.14$~Hz to $0.34$~Hz over a period of $\sim~20$ days. As for the recent 
GX 339-4 outburst, the shock also receded with a constant acceleration in two different ways, during the 
initial $4.2$ days slowly and then with a rapid acceleration, the only difference being that during the 
declining phase of the $2005$ GRO J1655-40 outburst, no QPOs were observed during the soft-intermediate state.

\subsection {Spectral energy distribution}

We carried out a spectral analysis to study the evolution of the spectral energy distribution during 
the outburst phase using ``standard 2" mode data from the RXTE Proportional Counter Unit 2 (PCU2) in 
the energy range of $2.5$ - $25$ keV. In general, the BH energy spectra (2-25 keV) were modeled by 
combining the two models {\it diskbb} and {\it powerlaw}. The first model was assumed to represent 
the thermal-component black body, and the later one the non-thermal component, which is 
mainly due to Comptonized photons. A Gaussian line of peak energy around $6.4$keV (iron-line emission) 
was used to obtain the best fit. For the spectral fitting, we kept the hydrogen column density (N$_{H}$) 
fixed at 5$\times$ 10$^{21}$ atoms cm$^{-2}$ (Mendez \& van der Klis 1997; Kong et al. 2000; 
Motta et al. 2009) and also assumed a $0.5$\% systematic error for the entire spectral fit. We did not 
include any smeared-edge model component (to account for a reflection component) in our spectral fitting.
In this {\it paper}, we present the results of the $146$ PCA observations spread over the outburst. 
The best-fit to the energy spectra were obtained by maintaining the value of the reduced $\chi^2$ 
($\chi^2$/DOF) of the fit at $\sim 1$. After obtaining the best-fit model spectra, we used the 
``error" command to calculate the error in the fitted parameters. All of these error values are of the 
$1\sigma$ confidence level. We used the ``flux LE HE err" command to calculate the errors in the flux 
for the energy range of LE and HE (in keV). 

The daily variations in the fitted model parameters of the $2.5 - 25$ keV RXTE/PCA spectra are plotted 
in Fig. 7, which clearly reveals the justification of separating the full outburst in the above-mentioned 
four states. The panels (a-d) are, respectively, the black body temperature $T_{in}$ in keV, the black body 
normalization factor, the photon index $\Gamma$, and the power-law normalization. Similarly in Fig. 8, the 
daily variations in the black body (BB) flux (panel a), the power-law (PL) flux (panel b), and the total flux 
(panel c) are shown. In panel (d), we show how the ratio of the black body to total flux changes daily.  
The convolution model `cflux' was used to calculate the flux contributions for the thermal (diskbb)
and non-thermal (power-law) components for the energy range of 2.5 - 25 keV.

\begin{figure}%[h]
\vbox{
\vskip 0.0cm
\centerline{
\includegraphics[scale=0.6,angle=0,width=8truecm]{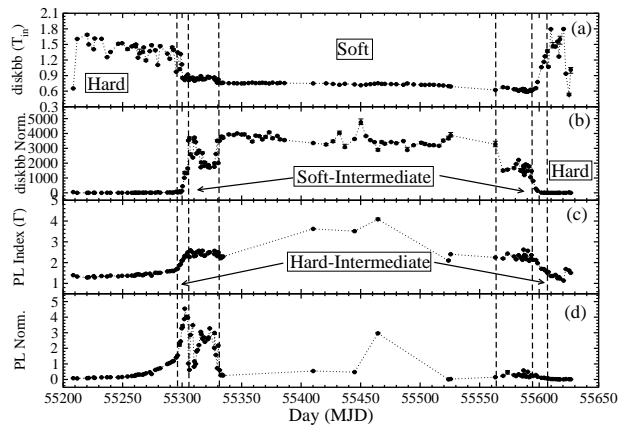}}
\caption{Fitted model parameters of RXTE $2.5 - 25$ keV PCA spectra plotted with time (MJD). 
In the panels we show variations of (a) disk black-body temperature (T$_{in}$) in keV, 
(b) disk black-body normalization, (c) power-law (PL) photon index ($\Gamma$), and 
(d) power-law normalization with time (day). The error bars in the model-fitted components 
are at $1~\sigma$ level.}}
\label{kn : fig7}
\end{figure}

\begin{table}
%\begin{table}[h]
\addtolength{\tabcolsep}{-4.50pt}
\vskip -0.2cm
\scriptsize
\centering
\centerline {Table 2}
\centerline {PCA Spectral model fitted parameters}
\vskip 0.2cm
\begin{tabular}{lcccccccc}
\hline
\hline
%%%   &   &   &    &  PCA    &    PCA     &    PCA    &      &      &   \\
Obs.&diskbb&diskbb&PL&PL&BB&PL&$\chi^2/DOF$\\
    &T$_{in}(keV)$&Norm.&Index($\Gamma$)&Norm.&Flux&Flux&\\
    (1)    &  (2)  & (3)  & (4)&   (5)   &  (6)    &    (7)      &  (8)    \\
\hline
26 &$1.304_{-0.031}^{+0.041}$&$17.66_{-0.840}^{+1.580}$&$1.549_{-0.016}^{+0.016}$&$0.808_{-0.019}^{+0.015}$&$0.524_{-0.049}^{+0.045}$&$10.36_{-0.066}^{+0.065}$&76.52/60 \\
35 &$1.012_{-0.041}^{+0.044}$&$207.9_{-8.700}^{+5.500}$&$2.024_{-0.062}^{+0.062}$&$3.062_{-0.156}^{+0.087}$&$1.651_{-0.127}^{+0.118}$&$10.18_{-0.316}^{+0.306}$&77.59/60 \\
51 &$0.878_{-0.009}^{+0.009}$&$ 1820_{-37.00}^{+35.00}$&$2.425_{-0.022}^{+0.022}$&$3.139_{-0.173}^{+0.184}$&$6.551_{-0.041}^{+0.041}$&$5.547_{-0.036}^{+0.036}$&95.26/61 \\
141&$0.735_{-0.005}^{+0.007}$&$ 3557_{-95.00}^{+101.0}$&$2.444_{-0.041}^{+0.025}$&$0.034_{-0.005}^{+0.002}$&$4.462_{-0.019}^{+0.019}$&$0.048_{-0.022}^{+0.015}$&14.43/28 \\
208&$0.663_{-0.016}^{+0.015}$&$ 1081_{-49.00}^{+58.00}$&$2.344_{-0.085}^{+0.085}$&$0.343_{-0.026}^{+0.021}$&$0.706_{-0.016}^{+0.016}$&$0.719_{-0.013}^{+0.013}$&80.42/61 \\
222&$1.065_{-0.046}^{+0.037}$&$15.99_{-1.940}^{+0.280}$&$1.661_{-0.032}^{+0.044}$&$0.096_{-0.006}^{+0.005}$&$0.166_{-0.011}^{+0.010}$&$0.966_{-0.008}^{+0.008}$&54.67/61 \\
226&$1.100_{-0.011}^{+0.034}$&$4.040_{-0.197}^{+0.254}$&$1.462_{-0.024}^{+0.023}$&$0.049_{-0.009}^{+0.010}$&$0.049_{-0.010}^{+0.009}$&$0.829_{-0.009}^{+0.009}$&42.47/62 \\
%1 &$1.466_{-0.046}^{+0.036}$&$13.68_{-0.790}^{+0.350}$&$1.508_{-0.027}^{+0.014}$&$0.719_{-0.024}^{+0.013}$&$4.275_{-0.015}^{+0.009}$&$4.159_{-0.012}^{+0.008}$&46.85/43 \\
%2 &$0.992_{-0.019}^{+0.029}$&$215.6_{-9.300}^{+7.900}$&$2.087_{-0.012}^{+0.005}$&$3.285_{-0.029}^{+0.013}$&$7.860_{-0.027}^{+0.011}$&$3.792_{-0.013}^{+0.010}$&44.93/44 \\
%3 &$0.881_{-0.008}^{+0.006}$&$1808_{-41.00}^{+24.00}$&$2.414_{-0.005}^{+0.005}$&$3.051_{-0.011}^{+0.008}$&$9.862_{-0.013}^{+0.009}$&$1.444_{-0.007}^{+0.006}$&72.46/42 \\
%4 &$0.735_{-0.003}^{+0.002}$&$3557_{-68.00}^{+56.00}$&$2.444_{-0.004}^{+0.001}$&$0.034_{-0.002}^{+0.001}$&$4.247_{-0.007}^{+0.005}$&$0.009_{-0.001}^{+0.001}$&13.99/26 \\
%5 &$0.656_{-0.013}^{+0.009}$&$1212_{-104.0}^{+89.00}$&$2.262_{-0.020}^{+0.013}$&$0.277_{-0.010}^{+0.006}$&$1.137_{-0.006}^{+0.004}$&$0.197_{-0.004}^{+0.002}$&52.00/44 \\
%6 &$1.061_{-0.010}^{+0.005}$&$15.58_{-0.560}^{+0.300}$&$1.703_{-0.011}^{+0.007}$&$0.107_{-0.005}^{+0.004}$&$0.551_{-0.004}^{+0.002}$&$0.358_{-0.010}^{+0.004}$&47.77/44 \\
%7 &$1.067_{-0.021}^{+0.007}$&$3.404_{-0.013}^{+0.008}$&$1.548_{-0.022}^{+0.006}$&$0.061_{-0.003}^{+0.002}$&$0.319_{-0.013}^{+0.004}$&$0.315_{-0.006}^{+0.003}$&23.14/44 \\
\hline
\end{tabular}
\noindent{
\leftline { PCA spectral fluxes are calculated in 2.5-40 keV energy range (except 2.5-17 keV for soft }
\leftline { state (Obs.=141) spectrum) and photon fluxes in units of $10^{-9}~ergs~cm^{-2}~s^{-1}$.}}
\end{table}

We note that during the soft spectral state, most of the spectra were fitted up to $10$ keV, because 
of the low and insignificant flux contribution above $>\sim 10$~keV (although some of the soft state 
spectra extends up to $\sim 15$ keV). From the nature of the variations in the power-law indices and 
disk black-body components, the full outburst is classified in four spectral states with a sequence of 
{\it hard $\rightarrow$ hard-intermediate $\rightarrow$ soft-intermediate $\rightarrow$ soft 
$\rightarrow$ soft-intermediate $\rightarrow$ hard-intermediate $\rightarrow$ hard} states (see,
also Fig. 2).

\begin{figure}
\vbox{
\vskip 0.5cm
\centerline{
\includegraphics[scale=0.6,angle=0,width=8truecm]{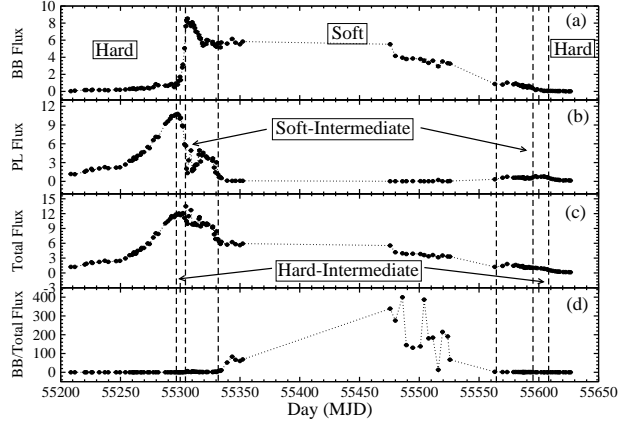}}
\caption {Derived properties of the daily flux variation. The panels are: (a) 2.5 - 25 keV 
bolometric flux due to the disk black-body component, termed black-body (BB) flux, 
(b) 2.5 - 25 keV flux due to the power-law component, termed power-law (PL) flux, 
(c) 2.5 - 25 keV total flux, and (d) the ratio of the total to PL fluxes. 
The error bars of the model fitted fluxes are at $1 \sigma$ level.}}
\label{kn : fig8}
\end{figure}

We also analyzed the PCA spectral data up to 40 keV (since the HEXTE spectral data were not useful 
for such high-energy spectral studies) in a few observations to search for the reflection features, 
high energy contributions, and particularly the high-energy power-law cutoff in the low-hard states. 
Hence, we fitted the energy spectra with a phenomenological model (diskbb + power-law), in addition 
to different combinations of models, such as `smedge', `reflect', and `cutoffpl'. We found that 
hard-state spectra were well-fitted by a phenomenological model without any extra model component, 
whereas the hard-intermediate state spectra could be fitted with a cut-off power-law ({\it cutoffpl}) 
modified by a reflection component ({\it reflect}) instead of a simple power-law model. An evolution 
of the high-energy cut-off in the hard states of GX 339-4 was previously reported for the 2006/2007 
outburst (Motta et al. 2009). On the other hand, the soft-intermediate state spectra were fitted with 
the same model combination along with a weak signature of the `smedge' component within the energy 
range from $\sim 7$ to $\sim 9$ keV, whereas soft-state spectra are well-fitted by a `diskbb' model 
and a weak power-law component (up to 15 keV).

In Figs. 9 (a-d), we plot the model-fitted energy spectra (up to 40 keV, except for the soft state 
up to $17$ keV), taken from four different spectral states (see, Fig. 2 labeled a, b, c, and d) of 
the present outburst of GX 339-4. A detailed modeling of the high energy spectrum in terms of a 
two component advective flow (TCAF) in different states is beyond the scope of the present work, 
and will be presented elsewhere. In each spectral plot (top panel), we show the unfolded energy 
spectrum with its individual model components and in the bottom panel, we show the variation in the 
ratio (data/model) over the fitted energy range. The best-fit model parameters for different states 
of the outburst are presented in Table 2.

\begin{figure}
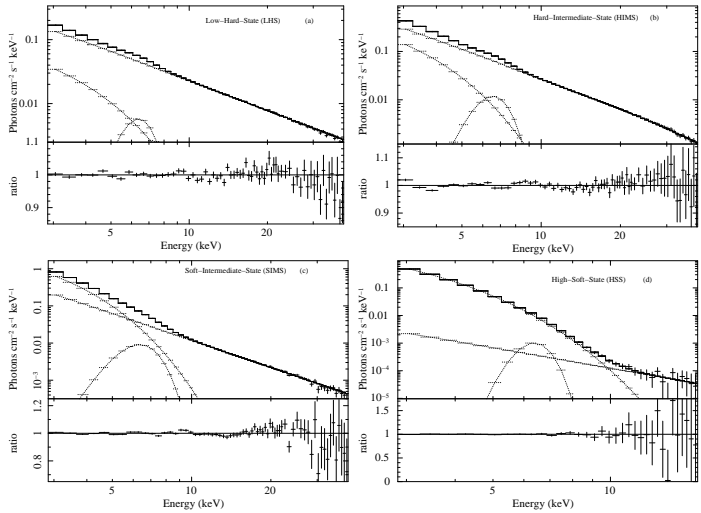

\vskip -0.0cm
\centerline{
       \includegraphics[scale=0.6,angle=270,width=4.5truecm]{fig9a.ps}\hskip 0.1cm
       \includegraphics[scale=0.6,angle=270,width=4.5truecm]{fig9b.ps}\\ }
\centerline{
       \includegraphics[scale=0.6,angle=270,width=4.5truecm]{fig9c.ps}\hskip 0.1cm
       \includegraphics[scale=0.6,angle=270,width=4.5truecm]{fig9d.ps}\\ }
\vspace{0.0cm}
\caption {(a-d): $2.5 - 40$ keV (except in Fig. d, where $2.5 - 17$ keV) RXTE/PCA model fitted 
spectra with various components from initial four different spectral states are shown. 
In top left panel (a) 2010 March 26 (Obs. Id: 95409-01-12-00), a typical hard state spectrum; 
in top right panel (b) hard-intermediate state spectrum of 2010 April 13 (Obs. Id: 95409-01-14-06); 
in bottom left (c) 2010 April 29 (Obs. Id: 95409-01-16-05), a typical soft-intermediate state 
spectrum, and in bottom right panel (d) 2010 September 19 (Obs. Id: 95409-01-35-02), soft state 
spectrum respectively are shown.}
\label{kn : fig9(a-d)}
\end{figure}

\subsection {Physical modeling of q-diagram with a two-component flow}

The HID of the outburst profile of GX 339-4 can be explained by a two-component advective flow 
(TCAF) model. An outburst is a time-dependent phenomenon, hence it can be analyzed completely 
by a steady-state model. We followed the prescription given in Mandal \& Chakrabarti (2010) to 
fit the main features of the HID of GX 339-4. The outburst might have been triggered by a sudden 
change in the viscosity that caused a conversion of sub-Keplerian into Keplerian matter. This 
increase in the Keplerian rate, along with the inward movement of the shock front (as suggested 
by the evolution of the QPOs) explains the outburst profile. In the rising phase, a high sub-Keplerian
rate makes the spectrum hard and a smooth increase in Keplerian rate increases the total
photon counts. The source GX 339-4 moved to an intermediate/soft state when the Keplerian rate 
was sufficiently high and the sub-Keplerian rate decreased to a low value such that the supply 
of soft photons was sufficient to cool the hot electrons in the post-shock region. 
Finally, in the decline phase, the Keplerian rate decreased and the sub-Keplerian rate approached 
the value at the quiescent state before the spectral state became hard again. In Fig. 10, the solid 
line with points shows the observed data, while the dotted line represents the model fit. 
We started the fitting process from the day that the QPO first appeared (22 March, 2010). 
In the rising phase ($t_r$), the Keplerian rate increased with power-law index ($\alpha_d$), 
the shock started to propagate inward with a constant velocity ($v_0$), and sub-Keplerian rate 
remains almost constant ($B_f$). The sub-Keplerian matter decreased with a power-law index 
($\alpha_{h1}$) between ($t_r$) and ($t_h$) and then started to increase. While the Keplerian 
rate decreased very slowly (linearly with a slope $0.0035$ Eddington rate/day) after $t_d$ 
until $t_h$, and then as a power-law ($\alpha_d$) with time. The shock started to recede 
outward after $t_s$. The slow decrease in Keplerian rate (from $t_d$ to $t_h$) kept the source 
in a very soft state for over $217$ days. This behavior of GX 339-4 is surprisingly different 
from GRO J1655-40 where Keplerian rate decreases by a constant power-law. The fit between 
`$C-D$' in Fig. 10 differs from the observed data since the result is very sensitive to the 
time variation in the parameters caused by the large number of soft photons in comparison with 
the hot electrons. In the decline phase, the sub-Keplerian rate in our model increases following 
a single power-law index but one may need to vary the power-law index to fit the data in the 
interval `$F-H$'. The parameters for the present fit are given in Table 3. The most important 
point is that the accretion-rate variation does not retrace its path when returning to quiescence, 
owing to the hysteresis behavior. For more detailed relations among the parameters, we refer to 
Mandal \& Chakrabarti (2010).

\begin{figure}
\vbox{
\vskip 0.0cm
\centerline{
\includegraphics[scale=0.6,angle=0,width=8truecm]{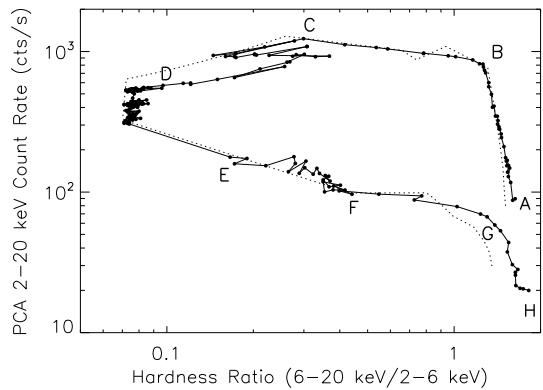}}
\vspace{0.0cm}
\caption{Two component accretion flow model fitted (dotted curve) HID of GX 339-4 
during the 2010-11 outburst. The solid curve with points show the observed data as 
shown in Fig. 2 and indicators $A - H$ are also have the same meanings.}}
\label{kn : fig10}
\end{figure}

\section {Discussion and concluding remarks}

As a very enigmatic transient X-ray BH candidate, GX 339-4 has underwent several X-ray outbursts 
after its discovery in the early 1970s. The detailed timing and spectral analysis results using 
RXTE/PCA instrument that we have presented in this paper reveal several very important aspects about 
the nature of the transient accretion process around a black hole. From the nature of the variations 
in the light curves, color-color variations, power spectra, energy spectra, and most importantly the 
variation in the QPO frequency, one can develop a comprehensive picture of what might be happening 
when such an outburst takes place.

The most natural assumption about the cause of the outburst is the change in the physical properties 
of the matter such as viscosity, perhaps owing to the enhanced magnetic activity. During the rising phase 
of the outburst, the viscosity may cause an increase in the accretion rate of the Keplerian matter. 
The outburst phase then subsides with the drop in viscosity and the Keplerian rate is reduced and 
the disk itself recedes. According to the two-component advective flow (TCAF) model (Chakrabarti 1990, 
Chakrabarti \& Titarchuk 1995), the initial phase is dominated by a low angular-momentum sub-Keplerian 
flow and the spectrum is found to be in a hard/low-hard state. As the outburst progresses, and the 
accretion rate in the Keplerian component increases and the object enters a soft spectral state. 
Finally, during the declining phase of the outburst, the Keplerian flow recedes while the 
sub-Keplerian matter rate remains roughly the same, which makes the spectrum harder. 
Debnath et al. (2008) found that this picture explains the 2005 GRO J1655-40 outburst quite 
successfully. From Figs. 7 and 8, one can also see that the above-mentioned picture is relevant 
to the recent GX 339-4 outburst, although the rate variation is not time-symmetric with respect 
to the peak. This causes the hysteresis loop behavior or the so-called `q'-diagram.

\begin{table}
\vskip -0.2cm
\addtolength{\tabcolsep}{-4.50pt}
\scriptsize
\centering
\centerline {Table 3}
\centerline {Parameters for Fig. 10}
%\vskip 0.2cm
\begin{tabular}{lccccccccc}
\hline
\hline
$t_r$ & $t_d$ & $t_h$ & $t_s$ & $\alpha_{d}$ & $\alpha_{h1}$ & $Xs_o$ & $Xs_i$ & $v_0$ & $B_f$\\
\cline{1-4}
&  Time(day) & & & & & ($r_g$)& ($r_g$)&($r_g$/day) & \\
\hline
19.0&30.0&247.0&325.0&-1.0&-0.6&400&14.0&12.8&2.0\\
\hline
\end{tabular}
\end{table}

It is clear from Fig. 1(a) that during the hard and hard-intermediate states of the rising phase 
of the outburst, $2-20$~keV photon count rate increases initially slowly (in the hard state), 
then rapidly (in the hard-intermediate state). In addition, from the hardness ratio plot of Fig. 1, 
one can note that the hardness ratios of the $6-20$ to $2-6$ keV photon rates initially decrease 
slowly (in the hard state) and then rapidly (in the hard-intermediate state). During both of these 
states, the low-energy disk black-body flux increases, whereas the high-energy power-law flux 
increases slowly 
in the hard state and rapidly decreases during the hard-intermediate state. During the hard state, 
the disk temperature ($T_{in}$) is $\sim 1.5$~keV, whereas during the hard-intermediate state
its value decreases to $\sim 0.9$~keV (see Fig. 7) within $\sim 9$ days. This is because during 
the hard spectral states both the high-energy and low-energy flux increase at almost the same rate 
(see, Fig. 8), whereas during the hard-intermediate state the high energy flux decreases owing to 
the decrease in the sub-Keplerian rate but the Keplerian rate remains almost constant. As a result, 
the disk becomes cool and its temperature decreases sharply. Furthermore, during the hard spectral 
state, the power-law photon index ($\Gamma$) increases slowly from $\sim 1.3$ to $\sim 1.7$, within 
a period of 
about three months, whereas during the hard-intermediate state, which has a period of $\sim 9$ days, 
it rapidly increases to $\sim 2.3$. This is because during the hard state, the spectra are dominated 
by the sub-Keplerian flow. As the day progresses, the Keplerian rate increases slowly in the 
hard state and rapidly in the hard-intermediate state, owing to the low-angular momentum 
sub-Keplerian flow moving on an infall timescale as the Keplerian flow moves towards the black 
hole on a viscous timescale.

The number of soft photons becomes comparable to those of hot electrons during the soft-intermediate 
spectral state. As a result, during this state the photon index became constant with $\Gamma \sim 2.4$. 
The spectra were then subsequently dominated by the Keplerian flow with a decrease in the sub-Keplerian 
flow and the spectrum becoming softer. As a result, the photon index increases further, becoming $> 3.0$. 
During the soft spectral state, the hard X-ray photon flux above $\sim 15$~keV, decreased rapidly, i.e., 
we can conclude that during this phase of the outburst, the inverse Comptonized sub-Keplerian flow decreases 
rapidly and the spectra are dominated by thermal Keplerian cool photons. Hence during the soft-state, the 
disk temperature ($T_{in}$) was observed to be $< 0.7$~keV, whereas in the previous soft-intermediate 
spectral state it had been observed to be constant at $\sim 0.9$~keV.

After spending about seven months in the Keplerian-flow-dominated soft spectral state, the source 
returned to a soft-intermediate spectral state. During this declining phase of the outburst, nearly 
identical behavior was observed as in the rising soft-intermediate spectral state. During this 
spectral state, $\Gamma$ was observed to be $\sim 2.3 - 2.5$ with $T_{in}$ at $\sim 0.6$~keV, owing 
to the slow increase of the sub-Keplerian component. The source then moved to a declining 
hard-intermediate spectral state with rapidly decreasing $\Gamma$ from $2.3$ (of the previous state) 
to $\sim 1.6$ and increasing $T_{in}$ from the previous state value of $\sim 0.6$~keV to 
$\sim 1.27$~keV within a few days. The physical reason 
for this is the deficiency of the fresh supply of Keplerian matter as the viscosity is reduced. Thus, on the 
whole, the fraction of Keplerian cold matter decreases, whereas the sub-Keplerian matter increases making the 
disk become hotter and causing the source to move to a hard spectral state. Here the power-law photon index 
$\Gamma$ is observed to decrease to $\sim 1.3$ from its previous state value of $\sim 1.6$ and its disk 
black-body temperature $T_{in}$ is increased to $\sim 1.8$, from its previous state value of $\sim 1.27$~keV.

During the rising phase of the outburst, strong QPOs were observed and in both the hard and hard-intermediate 
states the QPO frequencies increased monotonically with time. The formation of strong QPOs and the smooth 
variation in QPO frequencies during this phase (Debnath et al. 2010) indicates that the physical processes 
behind the formation of QPOs are identical each day and related to the dynamics of the infalling matter. 
We decided to choose the shock oscillation model (SOM) solution inside a sub-Keplerian disk, which has been 
demonstrated to have a stable oscillation for many dynamical timescales (Molteni, Sponholz \& Chakrabarti 1996; 
Ryu, Chakrabarti \& Molteni, 1997; Chakrabarti \& Manickam, 2000; Chakrabarti, Acharyya \& Molteni, 2004, 
Chakrabarti et al. 2005, 2008, 2009, Debnath et al. 2010, 2012). In this SOM solution, at the rising phase 
of the outburst, a shock wave moves toward the black hole, which oscillates either because of a resonance 
(where the cooling time of the flow is approximately the infall time; Molteni, Sponholz \& Chakrabarti, 1996) 
or because the Rankine-Hugoniot condition is not satisfied (Ryu, Chakrabarti \& Molteni, 1997) to form a 
steady shock. In the propagating oscillatory shock (POS) model, which is the propagatory case of SOM, it 
is easy to verify that the QPO frequencies (which are the inverses of the infall times from the post-shock 
to the BH) are simply related to the drifting of the shock towards (rising phase) or away from (declining phase) 
the BH. Debnath et al. (2010) inferred the movement of the shock wave towards BH for the same GX 339-4 
outburst with a constant velocity of $\sim 10~m~s^{-1}$. Similar movements of shock waves ($\sim 20~m~s^{-1}$) 
were observed by Chakrabarti et al. (2008, 2009) during the rising phases of GRO J1655-40 and XTE J1550-564. 
In the declining phase, Chakrabarti et al. (2008, 2009) observed that the shock waves receded from the BH. 
During the declining soft-intermediate spectral state of the outburst, QPOs are observed sporadically 
on and off at a frequency $\sim 2$ Hz for around $\sim 29$ days. After that, during the declining 
hard-intermediate spectral state, the QPO frequencies are observed to decrease monotonically from 
$6.42$~Hz to $1.149$Hz within a period of $\sim 10$ days. The QPO evolution during this phase is also 
fitted by the same POS model solution, the evolutions being more or less same as that of the declining 
phase of 2005 GRO J1655-40 outburst (Chakrabarti et al. 2008), in which the evolution of the decline 
phase had two parts. Then in the initial $3.5$ days, the shock moved away from the BH slowly and 
then rapidly. Here we also discern two behaviors. Initial $4.2$ days, the shock is found to move 
slowly (with an 
acceleration of $20~cm~sec^{-1}~day^{-1}$). Subsequently, it moves with a higher rate of 
acceleration ($175~cm~sec^{-1}~day^{-1}$). During this period, the shock velocity varies from 
$205~cm~sec^{-1}$ to $1785~cm~sec^{-1}$ with an intermediate value of $288~cm~sec^{-1}$ on the $4.2$nd day. At the same time, the shock is also found to move away from the BH from $84~r_g$ to $751~r_g$ 
passing via an intermediate location of $155~r_g$ on the transition ($4.2$nd) day. The universality 
of the behavior leads us to believe that the same processes are in operation in all the outbursting 
sources. However, an understanding of the details, such as the cause of the break in the QPO 
evolution during the declining phase, is still missing.

We have been able to model the evolution of power spectra in different spectral states with different 
combinations of Lorentzian and power-law like profiles, hence found that it can be explained with the 
diffusive propagation of perturbation in two different accretion flows (CT95, TSA07). The observed 
rms power decreases as the source moves from a hard-state to a soft-state, which is also quite natural 
in most outburst sources. The evolution pattern (i.e., q-diagram), which is characteristic of the 
outburst sources, of the 2010-11 outburst of GX 339-4 is modeled and explained with the two component 
advective flow (TCAF) model very successfully. 

The interpretation of the disk-jet dynamics in outburst BH sources has proven to be another 
challenging task. Several quasi-simultaneous observations of GX 339-4 in multi-wavelength studies 
(Homan et al. 2005;
Markoff et al. 2005; Maitra et al. 2009; Coriat et al. 2009) strongly suggest that the radio jets are 
associated with the hard state of the source and also illustrates the compact nature of the jet. 
The broadband continuum (radio through to X-rays) of GX 339-4 was described by a jet-dominated
model (see, Markoff et al. 2005; Maitra et al. 2009), in which the base of the jet was considered 
to be the hard X-ray emitting region. In reality, the base is nothing but the post-shock region, 
known as the CENBOL (i.e., the CENtrifugal pressure-dominated BOundary Layer) of TCAF, as the base 
of the jet, is able to explain the broadband nature of the spectrum. The broadband spectra of M87 
(Mandal \& Chakrabarti 2008) has been explained with TCAF model and the modeling of the 
multi-wavelength nature of the spectrum of GX 399-4 will be presented elsewhere. Since the basic 
cause of the outburst is believed to be the enhancement of viscous processes, simultaneous radio 
observations or the observations of the companion during the onset could provide valuable information 
about the outbursting properties. ASTROSAT, India's upcoming multi-wavelength Satellite, will 
possibly shed light on the `basic cause' of the outburst, as it carries instruments to study both 
the ultraviolet (observing the outer parts of the disk) and X-ray (observation of the inner parts 
of the  disk) emission from the outbursting system.

\section*{Acknowledgments}

We thank the anonymous referee for the detailed comments and suggestions to improve the 
quality of the manuscript.

%%%%%%%%%%%%%%%%%
%%%%%%%%%%%%%%%%%
%\end{document}
%%%%%%%%%%%%%%%%%
%%%%%%%%%%%%%%%%%
%\documentstyle[12pt]{article}
%\textwidth 16.0cm
%\textheight 21.0cm
%%\topmargin -0.8cm
%\hoffset=-1.2cm
%%\pagestyle{empty}
%\begin{document}
%%%%%%%%%%%%%%%%%%%%%%

\newpage
\vbox{
\vskip 10cm
\centering{\Large \bf Online Material}}

%\begin{table}[h]
\begin{table}
\vskip 0.5cm
%\addtolength{\tabcolsep}{-5.55pt}
\scriptsize
\centering
%\centering{\large \bf Online Material}
%\vskip 0.2cm
\centerline {Table A.1}
%\centerline {PCA Count rates, Hardness ratios and QPOs}
\centerline {PCA Count rates and Hardness ratios}
\vskip 0.2cm
\begin{tabular}{lccccccccc}
\hline
\hline
%  &   &   &  &  PCA  &  PCA  &    &  PCA  &  PCA  &    &     \\
Obs.&Id. &UT Date& MJD &2-6 keV&6-20 keV&$HR^a$&2-4 keV&4-30 keV&$HR^b$\\ %&QPO Freq.\\
 (1)  &(2)  &(3) & (4) & (5) & (6) & (7)  & (8)  &  (9)  & (10)  \\ %& (11) \\
\hline
1 &X-01-00&12/01/2010&55208.4918& 34.068& 55.570&1.631& 11.567& 87.874&7.597 \\  
2 &X-02-00&15/01/2010&55211.8234& 33.641& 53.838&1.600& 11.309& 85.405&7.552 \\
3 &X-03-00&24/01/2010&55220.1842& 45.508& 71.607&1.574& 15.342&112.722&7.347 \\
4 &X-03-04&25/01/2010&55221.5674& 50.625& 78.236&1.545& 17.062&123.154&7.218 \\
5 &X-04-00&29/01/2010&55225.7432& 57.902& 90.411&1.561& 19.495&141.606&7.264 \\
6 &X-04-04&30/01/2010&55226.5176& 54.635& 83.877&1.535& 18.446&131.985&7.155 \\
\hline                                                                       
\end{tabular}
\noindent{
\leftline {$X$ means initial 95409-01 part of the observation Ids, $^a$ and $^b$ are the PCA count rate ratios 
between 6-20 keV \& 2-6 keV }
\leftline {and between 4-30 keV \& 2-4 keV respectively.}}
\end{table}

%\begin{table}[h]
\begin{table}
\vskip 0.5cm
%\addtolength{\tabcolsep}{-5.55pt}
\scriptsize
\centering
\centerline {Table A.1 (Cont'd)}
%\centerline {PCA Count rates, Hardness ratios and QPOs}
\centerline {PCA Count rates and Hardness ratios}
\vskip 0.2cm
\begin{tabular}{lccccccccc}
\hline
\hline
%  &   &   &  &  PCA  &  PCA  &    &  PCA  &  PCA  &    &     \\
Obs.&Id. &UT Date& MJD &2-6 keV&6-20 keV&$HR^a$&2-4 keV&4-30 keV&$HR^b$\\ %&QPO Freq.\\
 (1)  &(2)  &(3) & (4) & (5) & (6) & (7)  & (8)  &  (9)  & (10)  \\ %& (11) \\
\hline
 51&X-16-05&29/04/2010&55315.7128&830.042&255.000&0.307&398.985&697.461&1.748 \\
 52&X-17-00&30/04/2010&55316.1217&831.164&256.002&0.308&399.588&699.519&1.751 \\
 53&X-17-01&01/05/2010&55317.1015&765.662&174.112&0.227&385.237&563.398&1.462 \\
 54&X-17-02&02/05/2010&55318.4475&678.491&249.356&0.368&324.988&615.553&1.894 \\
 55&X-17-03&03/05/2010&55319.1331&691.092&227.232&0.329&335.814&594.513&1.770 \\
 56&X-17-05&05/05/2010&55321.7343&741.704&208.849&0.282&361.549&598.468&1.655 \\
\hline                                                                       
\end{tabular}
\noindent{
\leftline {$X$ means initial 95409-01 part of the observation Ids, $^a$ and $^b$ are the PCA count rate ratios
between 6-20 keV \& 2-6 keV }
\leftline {and between 4-30 keV \& 2-4 keV respectively.}}
\end{table}

%\begin{table}[h]
\begin{table}
\vskip 0.5cm
%\addtolength{\tabcolsep}{-5.55pt}
\scriptsize
\centering
\centerline {Table A.1 (Cont'd)}
%\centerline {PCA Count rates, Hardness ratios and QPOs}
\centerline {PCA Count rates and Hardness ratios}
\vskip 0.2cm
\begin{tabular}{lccccccccc}
\hline
\hline
%  &   &   &  &  PCA  &  PCA  &    &  PCA  &  PCA  &    &     \\
Obs.&Id. &UT Date& MJD &2-6 keV&6-20 keV&$HR^a$&2-4 keV&4-30 keV&$HR^b$\\ %&QPO Freq.\\
 (1)  &(2)  &(3) & (4) & (5) & (6) & (7)  & (8)  &  (9)  & (10) \\ % & (11) \\
\hline
101&X-25-00&25/06/2010&55372.4818&496.488& 37.380&0.075&277.234&261.475&0.943 \\
102&X-25-01&26/06/2010&55373.1282&495.948& 35.655&0.072&277.563&258.609&0.932 \\
103&X-25-02&27/06/2010&55374.4405&497.821& 37.460&0.075&278.797&261.427&0.938 \\
104&X-25-03&28/06/2010&55375.2216&499.664& 37.123&0.074&281.350&260.514&0.926 \\
105&X-25-04&28/06/2010&55376.0158&510.545& 39.937&0.078&283.824&271.326&0.956 \\
106&X-25-05&30/06/2010&55377.1699&501.921& 38.648&0.077&281.340&264.264&0.939 \\
\hline                                                                       
\end{tabular}
\noindent{
\leftline {$X$ means initial 95409-01 part of the observation Ids, $^a$ and $^b$ are the PCA count rate ratios
between 6-20 keV \& 2-6 keV }
\leftline {and between 4-30 keV \& 2-4 keV respectively.}}
\end{table}

%\begin{table}[h]
\begin{table}
\vskip 0.5cm
%\addtolength{\tabcolsep}{-5.55pt}
\scriptsize
\centering
\centerline {Table A.1 (Cont'd)}
%\centerline {PCA Count rates, Hardness ratios and QPOs}
\centerline {PCA Count rates and Hardness ratios}
\vskip 0.2cm
\begin{tabular}{lccccccccc}
\hline
\hline
%  &   &   &  &  PCA  &  PCA  &    &  PCA  &  PCA  &    &     \\
Obs.&Id. &UT Date& MJD &2-6 keV&6-20 keV&$HR^a$&2-4 keV&4-30 keV&$HR^b$\\ % &QPO Freq.\\
 (1)  &(2)  &(3) & (4) & (5) & (6) & (7)  & (8)  &  (9)  & (10)  \\ %& (11) \\
\hline
151&X-36-05&29/09/2010&55468.7747&377.649&29.941&0.079&215.638&197.448&0.916 \\
152&X-36-06&30/09/2010&55469.6164&373.284&29.358&0.079&212.293&195.856&0.923 \\
153&X-37-00&01/10/2010&55470.3209&377.974&29.140&0.077&215.969&196.153&0.908 \\
154&X-37-01&02/10/2010&55471.0500&381.720&29.518&0.077&216.943&199.166&0.918 \\
155&X-37-02&03/10/2010&55472.4220&381.244&29.455&0.077&217.186&198.148&0.912 \\
156&X-37-05&06/10/2010&55475.5642&379.672&28.999&0.076&216.787&197.314&0.910 \\
\hline                                                                                       
\end{tabular}                                                                                
\noindent{
\leftline {$X$ \& $Y$ indicate 95409-01 \& 96409-01 of the observation Ids respectively, $^a$ and $^b$ 
are the PCA count rate ratios between }
\leftline {6-20 keV \& 2-6 keV and between 4-30 keV \& 2-4 keV respectively.}}
\end{table}

%\begin{table}[h]
\begin{table}
\vskip 0.5cm
%\addtolength{\tabcolsep}{-5.55pt}
\scriptsize
\centering
\centerline {Table A.1 (Cont'd)}
%\centerline {PCA Count rates, Hardness ratios and QPOs}
\centerline {PCA Count rates and Hardness ratios}
\vskip 0.2cm
\begin{tabular}{lccccccccc}
\hline
\hline
%  &   &   &  &  PCA  &  PCA  &    &  PCA  &  PCA  &    &     \\
Obs.&Id. &UT Date& MJD &2-6 keV&6-20 keV&$HR^a$&2-4 keV&4-30 keV&$HR^b$\\ % &QPO Freq.\\
 (1)  &(2)  &(3) & (4) & (5) & (6) & (7)  & (8)  &  (9)  & (10)  \\ %& (11) \\
\hline
201&Y-02-03& 13/01/2011& 55574.8832&105.590&30.530&0.289& 61.287& 80.941&1.321 \\
202&Y-03-00& 15/01/2011& 55576.8682&114.455&34.537&0.302& 64.278& 90.558&1.409 \\
203&Y-03-01& 17/01/2011& 55578.8929&101.326&32.771&0.323& 56.928& 83.054&1.459 \\
204&Y-03-02& 19/01/2011& 55580.6278&110.706&36.851&0.333& 61.140& 92.614&1.515 \\
205&Y-04-00& 21/01/2011& 55582.7131&101.285&34.450&0.340& 56.508& 85.424&1.512 \\
206&Y-04-01& 23/01/2011& 55584.3696& 87.364&32.322&0.370& 48.136& 77.263&1.605 \\
\hline                                                                                       
\end{tabular}                                                                                
\noindent{
\leftline {$X$ \& $Y$ indicate 95409-01 \& 96409-01 of the observation Ids respectively, $^a$ and $^b$ 
are the PCA count rate ratios between }
\leftline {6-20 keV \& 2-6 keV and between 4-30 keV \& 2-4 keV respectively.}}
\end{table}

%%%%%%%%%%%%%%%%%%%%%%%%%%%%%%%%%%%%%%%%%%%%%%%%%%%%%%%%%%%%%%%%%%%
%\begin{table}[h]
\begin{table}
\vskip 4cm
\scriptsize
\centering
\centerline {Table A.2}
\centerline {Observed QPO fitted parameters (rising phase)}
\vskip 0.0cm
\begin{tabular}{lccccccc}
\hline
\hline
% & & & & & RMS \\
Obs. &Id   &UT Date&  MJD  &$\nu$  &$\Delta\nu$&  Q  &rms(\%) \\%&Type \\
 (1) & (2) &  (3)  &  (4)  &  (5)  &  (6)      & (7) & (8)   \\ %& (9) \\
\hline

  24&95409-01-11-02 &22/03/2010&55277.4793&$0.102_{-0.008}^{+0.006}$&$0.032_{-0.011}^{+0.022}$&3.207&17.438 \\ 
  25&95409-01-11-03 &24/03/2010&55279.5718&$0.113_{-0.005}^{+0.006}$&$0.045_{-0.001}^{+0.001}$&2.500&18.879 \\
  26&95409-01-12-00 &26/03/2010&55281.5923&$0.134_{-0.006}^{+0.006}$&$0.022_{-0.002}^{+0.001}$&6.091&10.841 \\
  27&95409-01-12-03 &01/04/2010&55287.6027&$0.202_{-0.004}^{+0.003}$&$0.062_{-0.006}^{+0.006}$&3.237&19.044 \\
  28&95409-01-13-00 &03/04/2010&55289.6289&$0.261_{-0.006}^{+0.004}$&$0.080_{-0.005}^{+0.004}$&3.254&17.158 \\
  29&95409-01-13-02 &05/04/2010&55291.6595&$0.317_{-0.003}^{+0.005}$&$0.085_{-0.003}^{+0.005}$&3.751&16.952 \\
\hline
\end{tabular}
\noindent{
\leftline {Here $\nu$, $\Delta \nu$ \& $Q$ (=$\nu/\Delta \nu$) are centroid frequency (in Hz), FWHM (in Hz) and 
coherence factor of QPOs. }
\leftline {N.B.: Only information about principal QPOs are tabulated here.}}
\end{table}

%\begin{table}[h]
\begin{table}
\scriptsize
\centering
\vskip 0.2cm
\centerline {Table A.2 (Cont'd)}
\centerline {Observed QPO fitted parameters (declining phase)}
\vskip 0.0cm
\begin{tabular}{lccccccc}
\hline
\hline
% & & & & & RMS \\
Obs. &Id   &UT Date&  MJD  &$\nu$  &$\Delta\nu$&  Q  &rms(\%)\\ %&Type \\
 (1) & (2) &  (3)  &  (4)  &  (5)  &  (6)      & (7) & (8)   \\ %& (9) \\
\hline
                                                                                                           
% 206&96409-01-04-01 &23/01/2011&55584.3696&$0.662_{-0.039}^{+0.042}$&$0.113_{-0.014}^{+0.017}$&5.868& 3.779 \\
% 206&96409-01-04-01 &23/01/2011&55584.3696&$1.065_{-0.055}^{+0.094}$&$0.129_{-0.020}^{+0.001}$&8.262& 3.039 \\
 206&96409-01-04-01 &23/01/2011&55584.3696&$1.722_{-0.077}^{+0.052}$&$0.202_{-0.003}^{+0.054}$&8.516& 4.470 \\
 207&96409-01-04-04 &24/01/2011&55585.9533&$2.049_{-0.157}^{+0.199}$&$0.878_{-0.447}^{+0.811}$&2.332& 8.363 \\
 208&96409-01-04-02 &25/01/2011&55586.5030&$2.136_{-0.160}^{+0.138}$&$1.065_{-0.068}^{+0.087}$&2.005& 9.725 \\
 216&96409-01-05-01 &30/01/2011&55591.6236&$1.705_{-0.117}^{+0.133}$&$0.661_{-0.098}^{+0.070}$&2.580& 8.105 \\
% 217&96409-01-05-05 &31/01/2011&55592.7394&$1.434_{-0.045}^{+0.107}$&$0.431_{-0.034}^{+0.083}$&3.325& 6.807 \\
 217&96409-01-05-05 &31/01/2011&55592.7394&$2.165_{-0.110}^{+0.223}$&$0.271_{-0.004}^{+0.005}$&7.986& 5.587 \\
 218&96409-01-05-02 &01/02/2011&55593.5129&$1.742_{-0.078}^{+0.056}$&$0.408_{-0.151}^{+0.211}$&4.268& 8.066 \\
\hline
\end{tabular}
\noindent{
\leftline {Here $\nu$, $\Delta \nu$ \& $Q$ (=$\nu/\Delta \nu$) are centroid frequency (in Hz), FWHM (in Hz) and 
coherence factor of QPOs.}
\leftline {N.B.: Only information about principal QPOs are tabulated here.}}
\end{table}

%%%%%%%%%%%%%%%%%%%%%%%%%%%%%%%%%%%%%%%%%%%%%%%%%%%%%%%%%%%%%%%%%%%

%\begin{table}[h]
\begin{table}
\vskip 0.5cm
\addtolength{\tabcolsep}{-3.00pt}
\scriptsize
\centering
\centerline {Table A.3}
\centerline {2.5-25 keV PCA Spectrum Fitted Parameters}
\vskip 0.2cm
\begin{tabular}{lcccccccccc}
\hline
\hline
%Obs.&Id&UT Date&MJD&diskbb&diskbb&Power-Law&Power-Law&Black Body&Power-Law&$\chi^2_{red.}$\\
Obs.&Id&UT&MJD&DBB&DBB&PL&PL&DBB&PL&$\chi^2/DOF$\\
   & & & &T$_{in}(keV)$&Norm.&Index($\Gamma$)&Norm.&Flux&Flux&\\
 (1)  &  (2)  & (3)  & (4)& (5) & (6) & (7) &  (8) & (9) & (10) & (11)  \\
\hline

  1&X-01-00 &12/01/10&55208.4916&$0.653_{-0.004}^{+0.003}$&$49.95_{-1.500}^{+0.500}$&$1.395_{-0.007}^{+0.004}$&$0.084_{-0.002}^{+0.001}$&$0.028_{-0.004}^{+0.005}$&$1.168_{-0.029}^{+0.028}$&29.38/43 \\
  2&X-02-00 &15/01/10&55211.8201&$1.605_{-0.007}^{+0.005}$&$1.073_{-0.008}^{+0.006}$&$1.329_{-0.005}^{+0.003}$&$0.069_{-0.003}^{+0.002}$&$0.087_{-0.007}^{+0.007}$&$1.129_{-0.024}^{+0.024}$&46.44/44 \\
  3&X-03-00 &24/01/10&55220.1839&$1.686_{-0.016}^{+0.007}$&$1.547_{-0.009}^{+0.004}$&$1.283_{-0.004}^{+0.002}$&$0.084_{-0.002}^{+0.001}$&$0.159_{-0.009}^{+0.010}$&$1.531_{-0.036}^{+0.035}$&32.90/44 \\
  4&X-03-04 &25/01/10&55221.5630&$1.500_{-0.004}^{+0.006}$&$2.704_{-0.007}^{+0.005}$&$1.298_{-0.011}^{+0.006}$&$0.097_{-0.003}^{+0.004}$&$0.160_{-0.009}^{+0.008}$&$1.706_{-0.026}^{+0.026}$&35.93/43 \\
  5&X-04-00 &29/01/10&55225.7256&$1.407_{-0.006}^{+0.005}$&$3.152_{-0.005}^{+0.004}$&$1.349_{-0.006}^{+0.005}$&$0.129_{-0.006}^{+0.003}$&$0.136_{-0.009}^{+0.009}$&$2.002_{-0.021}^{+0.021}$&29.03/43 \\
\hline
\end{tabular}
\noindent{
\leftline {Here X=95409-01 means the initial part of the observation Ids and DBB, PL represent spectrum fitted model components:}
\leftline {disk black body \& power-law respectively. Here DBB \& PL fluxes are calculated in units of 
$10^{-9}~ergs~cm^{-2}~s^{-1}$.}}
\end{table}

\begin{table}
\vskip 0.5cm
\addtolength{\tabcolsep}{-3.00pt}
\scriptsize
\centering
\centerline {Table A.3 (Cont'd)}
\centerline {2.5-25 keV PCA Spectrum Fitted Parameters}
\vskip 0.2cm
\begin{tabular}{lcccccccccc}
\hline
\hline
Obs.&Id&UT&MJD&DBB&DBB&PL&PL&DBB&PL&$\chi^2/DOF$\\
   & & & &T$_{in}(keV)$&Norm.&Index($\Gamma$)&Norm.&Flux&Flux&\\
 (1)  &  (2)  & (3)  & (4)& (5) & (6) & (7) &  (8) & (9) & (10) & (11)  \\
\hline

 51&X-16-05 &29/04/10&55315.7125&$0.878_{-0.009}^{+0.009}$&$ 1820_{-37.00}^{+35.00}$&$2.425_{-0.022}^{+0.022}$&$3.139_{-0.173}^{+0.184}$&$6.550_{-0.041}^{+0.041}$&$5.547_{-0.036}^{+0.036}$&95.26/61 \\
 52&X-17-00 &30/04/10&55316.1214&$0.865_{-0.006}^{+0.007}$&$ 2044_{-50.00}^{+40.00}$&$2.390_{-0.004}^{+0.003}$&$2.900_{-0.003}^{+0.006}$&$6.274_{-0.118}^{+0.116}$&$4.938_{-0.120}^{+0.117}$&47.48/44 \\
 53&X-17-01 &01/05/10&55317.1013&$0.827_{-0.020}^{+0.008}$&$ 2679_{-80.00}^{+30.00}$&$2.573_{-0.025}^{+0.028}$&$2.615_{-0.024}^{+0.013}$&$6.167_{-0.130}^{+0.127}$&$3.165_{-0.100}^{+0.097}$&39.48/43 \\
 54&X-17-02 &02/05/10&55318.4481&$0.833_{-0.013}^{+0.012}$&$ 1816_{-30.00}^{+20.00}$&$2.332_{-0.015}^{+0.014}$&$2.698_{-0.025}^{+0.005}$&$5.672_{-0.125}^{+0.124}$&$4.165_{-0.106}^{+0.104}$&51.58/45 \\
 55&X-17-03 &03/05/10&55319.1281&$0.832_{-0.017}^{+0.016}$&$ 2040_{-50.00}^{+40.00}$&$2.336_{-0.029}^{+0.009}$&$2.416_{-0.020}^{+0.012}$&$5.370_{-0.106}^{+0.104}$&$4.561_{-0.107}^{+0.105}$&45.86/44 \\
\hline
\end{tabular}
\noindent{
\leftline {Here X=95409-01 \& Y=96409-01 mean the initial part of the observation Ids and DBB, PL represent spectrum fitted model components:}
\leftline {disk black body \& power-law respectively. The unit of the spectral fluxes : $10^{-9}~ergs~cm^{-2}~s^{-1}$. 
$^*$ spectrum is fitted \& flux are in 2.5-17 keV.}}
\end{table}

\begin{table}
\vskip 0.5cm
\addtolength{\tabcolsep}{-3.00pt}
\scriptsize
\centering
\centerline {Table A.3 (Cont'd)}
\centerline {2.5-25 keV PCA Spectrum Fitted Parameters}
\vskip 0.2cm
\begin{tabular}{lcccccccccc}
\hline
\hline
Obs.&Id&UT&MJD&DBB&DBB&PL&PL&DBB&PL&$\chi^2/DOF$\\
   & & & &T$_{in}(keV)$&Norm.&Index($\Gamma$)&Norm.&Flux&Flux&\\
 (1)  &  (2)  & (3)  & (4)& (5) & (6) & (7) &  (8) & (9) & (10) & (11)  \\
\hline

211&Y-04-07 &26/01/11&55587.5035&$0.631_{-0.003}^{+0.003}$&$ 1198_{-74.00}^{+20.00}$&$2.338_{-0.009}^{+0.005}$&$0.302_{-0.004}^{+0.002}$&$0.565_{-0.017}^{+0.015}$&$0.567_{-0.028}^{+0.026}$&42.98/44 \\
212&Y-04-06G&26/01/11&55587.8115&$0.637_{-0.015}^{+0.009}$&$ 1364_{-68.00}^{+25.00}$&$2.363_{-0.039}^{+0.010}$&$0.301_{-0.004}^{+0.005}$&$0.535_{-0.015}^{+0.014}$&$0.538_{-0.028}^{+0.027}$&50.73/44 \\
213&Y-04-08 &27/01/11&55588.5548&$0.592_{-0.004}^{+0.003}$&$ 1909_{-20.00}^{+20.00}$&$2.326_{-0.014}^{+0.006}$&$0.229_{-0.005}^{+0.003}$&$0.565_{-0.014}^{+0.014}$&$0.440_{-0.023}^{+0.022}$&48.30/44 \\
214&Y-05-00 &28/01/11&55589.2045&$0.595_{-0.007}^{+0.006}$&$ 1886_{-90.00}^{+20.00}$&$2.168_{-0.012}^{+0.004}$&$0.184_{-0.005}^{+0.004}$&$0.595_{-0.017}^{+0.017}$&$0.483_{-0.019}^{+0.018}$&45.77/44 \\
215&Y-05-04 &29/01/11&55590.4420&$0.589_{-0.021}^{+0.012}$&$ 1491_{-19.00}^{+14.00}$&$2.569_{-0.025}^{+0.019}$&$0.493_{-0.019}^{+0.010}$&$0.442_{-0.032}^{+0.030}$&$0.601_{-0.026}^{+0.025}$&53.21/44 \\
\hline
\end{tabular}
\noindent{
\leftline {Here Y=96409-01 means the initial part of the observation Ids and DBB, PL represent spectrum fitted model components:}
\leftline {disk black body \& power-law respectively. The unit of the spectral fluxes : $10^{-9}~ergs~cm^{-2}~s^{-1}$.}}
\end{table}

%%%%%%%%%%%

\end{document}